\documentclass[aps,prb,twocolumn,superscriptaddress,amsmath,amssymb,longbibliography,]{revtex4-2}
\usepackage[colorlinks=true,linkcolor=blue,anchorcolor=red,citecolor=blue, urlcolor=blue]{hyperref}
\usepackage{graphicx} 
\usepackage{dcolumn} 
\usepackage{bm} 
\usepackage{color} 
\usepackage{xfrac}

\begin{document}

\title{Tunable second harmornic in altermagnetic Josephson junctions} 

\author{Hai-Peng Sun}
\email{hpsun@sztu.edu.cn}
\affiliation{Shenzhen Key Laboratory of Ultraintense Laser and Advanced Material Technology, Center for Intense Laser Application Technology, and College of Engineering Physics, Shenzhen Technology University, Shenzhen 518118, China}
\affiliation{Institute for Theoretical Physics and Astrophysics, University of W\"urzburg, 97074 W\"urzburg, Germany}
\affiliation{W\"urzburg-Dresden Cluster of Excellence ct.qmat, 97074 W\"urzburg, Germany}

\author{Song-Bo Zhang}
\email{songbozhang@ustc.edu.cn}
\affiliation{Hefei National Laboratory, Hefei, Anhui, 230088, China}
\affiliation{International Center for Quantum Design of Functional Materials (ICQD),  
University of Science and Technology of China, Hefei, Anhui 230026, China}
    
\author{Chang-An Li}
\affiliation{Institute for Theoretical Physics and Astrophysics, University of W\"urzburg, 97074 W\"urzburg, Germany}
\affiliation{W\"urzburg-Dresden Cluster of Excellence ct.qmat, 97074 W\"urzburg, Germany}

\author{Bj\"orn Trauzettel}
\affiliation{Institute for Theoretical Physics and Astrophysics, University of W\"urzburg, 97074 W\"urzburg, Germany}
\affiliation{W\"urzburg-Dresden Cluster of Excellence ct.qmat, 97074 W\"urzburg, Germany}

\date{\today }

\begin{abstract}
We study the influence of external electric and Zeeman fields on the Josephson effect in a planar superconductor/altermagnet/superconductor junction. 
Remarkably, we find that the current-phase relation (CPR) can be forward or backward skewed due to a pronounced second harmonic term. It decisively depends on the altermagnetic field strength. This second harmonic can be measured directly using double SQUID devices. The CPR skewness can be effectively manipulated by electric gating. Moreover, we identify two additional impacts of external electric and magnetic fields on the Josephson current: (i) Fields can induce 0-$\pi$ transitions. (ii) Fields can substantially enhance the critical current. This enhancement is surprising since supercurrents are typically suppressed by magnetic fields.
\end{abstract}

\maketitle

 {\color{blue}\emph{Introduction.}}---
%\section{Introduction}
Altermagnism is a newly discovered magnetic phase categorized by spin-group symmetries~\cite{Smejkal22PRXEmerging,Smejkal22PRXConventional,CJWu07PRB,Naka19NCSpin,shao2021spin,Hayami19JPSJMomentumDependent,Ahn19PRBAntiferromagnetism,Smejkal20SACrystal,Yuan20PRBGiant,Bai24Altermagnetism}. It is an unconventional collinear antiferromagnet, where opposite spins at different sublattices are related by crystal rotation or reflection symmetry~\cite{Smejkal22PRXEmerging,Smejkal22PRXConventional}, in contrast to classical antiferromagnets, where the spins are related via inversion or lattice translation symmetry. Hence, the combined time-reversal and translation (or inversion) symmetry is broken in altermagnets, leading to anisotropic spin-split energy bands in momentum space. 
This phenomenon has been predicted in various candidate materials~\cite{Berlijn17PRLItinerant,Ahn19PRBAntiferromagnetism,Smejkal20SACrystal,Hayami19JPSJMomentumDependent,Mazin21PNASPrediction,Smejkal22PRXConventional,Yuan20PRBGiant,Smejkal22PRXGiant,Smejkal22PRXEmerging,Mazin23PRBAltermagnetism,Guo23MTPSpinsplit,Lin24Observation}, such as 
RuO$_2$~\cite{Berlijn17PRLItinerant,Ahn19PRBAntiferromagnetism,Smejkal20SACrystal}, 
MnTe~\cite{Smejkal22PRXConventional,Mazin23PRBAltermagnetism}, CrSb~\cite{Smejkal22PRXConventional,Guo23MTPSpinsplit},FeSb$_2$~\cite{Mazin21PNASPrediction,Smejkal22PRXConventional}, Mn$_5$Si$_3$~\cite{Smejkal22PRXGiant}, and CrNb$_4$S$_8$~\cite{Guo23MTPSpinsplit}, and  confirmed experimentally in Mn$_5$Si$_3$~\cite{Reichlova24NCObservation}, MnTe~\cite{Krempasky24NAltermagnetic,Lee24PRLBroken,Osumi24PRBObservation,Orlova24JLCrossover}, and CrSb~\cite{Reimers24NCDirect,Ding24PRLLarge,Yang25NCThreedimensional}. 
One notable feature of altermagnets is spin-momentum locking, protected by spin groups~\cite{Smejkal22PRXConventional,Smejkal22PRXEmerging,Liu22PRXSpinGroup}. Similar spin-momentum locking can be achieved by twisting antiferromagnetic bilayers~\cite{He23PRLNonrelativistic,Sheoran24PRMNonrelativistic,Liu24PRLTwisted,Zeng24PRBBilayer}. In these systems, the altermagnetic field strength (AFS) depends crucially on the twist angle.

\begin{figure}[!htp] 
\centering
\includegraphics[width=0.48\textwidth]
{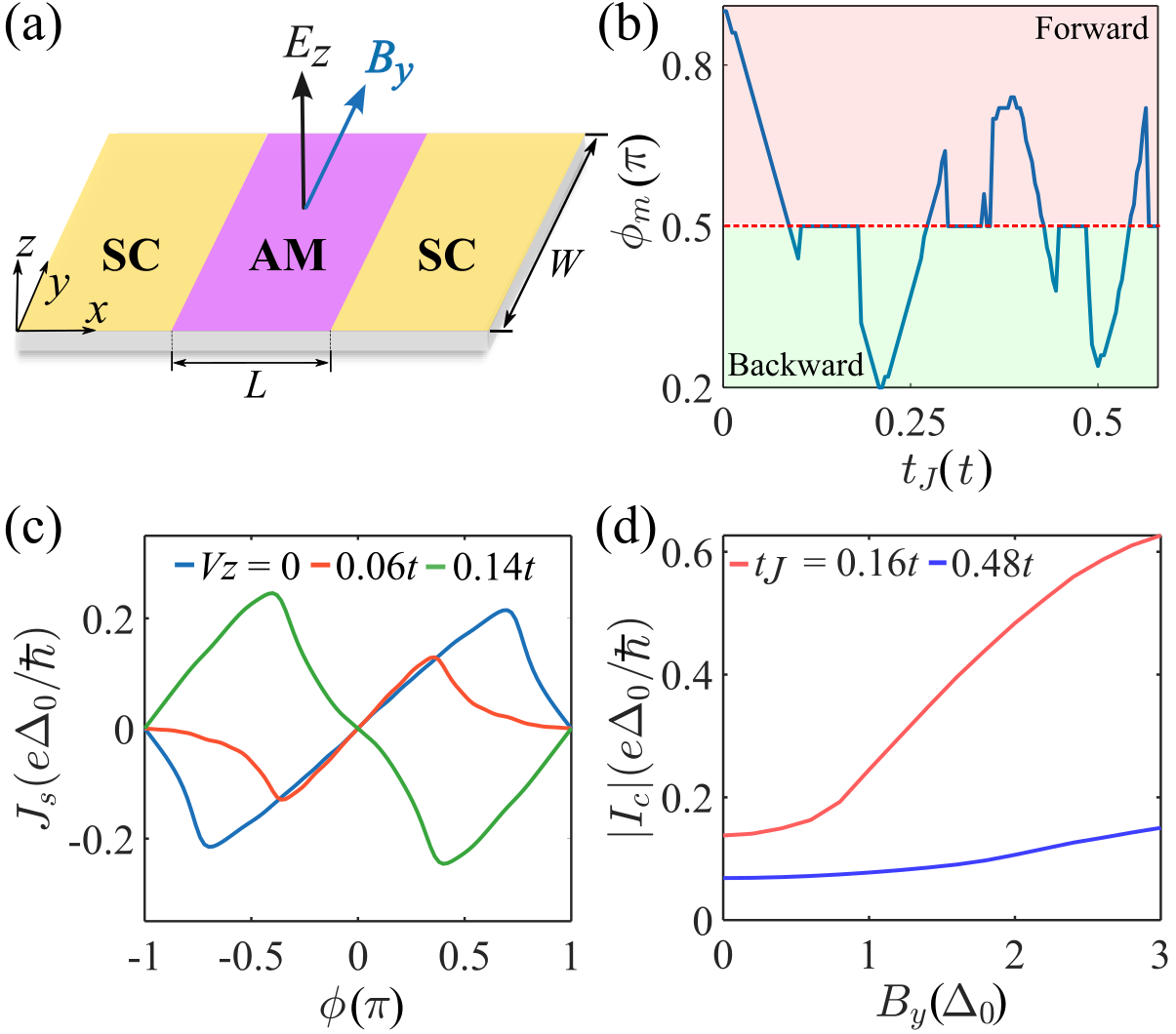}\\
\setlength{\abovecaptionskip}{-1pt}
\caption{(a) Schematic of the AMJJ under external electric ($E_z$) or magnetic ($B_y$) fields. Yellow and purple regions indicate the $s$-wave superconducting leads and altermagnet (AM), respectively.
Black and blue arrows indicate external fields applied to AM. (b) Skewness change of the CPR with increasing AFS $t_J$. $\phi_m$ is the phase difference of the maximum Josephson current. 
(c) CPRs of the AMJJ with $t_J=0.4t$ for various gate potentials $V_z$. Supercurrents are in units of $e\Delta_{0}/\hbar$.
(d) Absolute value of the critical current $I_c$ as a function of in-plane Zeeman field $B_y$ for different
$t_J$. 
}\label{Fig:setup}
\vspace{-2em}
\end{figure}

Recently, the interplay of altermagnetism and superconductivity has triggered intensive interest~\cite{Zhu2023Topological,Wei24PRBGapless,Sun23PRBAndreev,Papaj23PRBAndreev,Nagae25PRBSpinpolarized,Zhang24NCFinitemomentum,Ouassou23PRLDc,Beenakker23PRBPhaseshifted,Sumita2023Fulde,Chakraborty23Zero,Cheng2024Orientation},  offering unique opportunities for exploring novel phenomena and advanced applications in superconducting spintronics and quantum computing. In particular, orientation-dependent Andreev reflection has been investigated at superconductor/altermagnet interfaces~\cite{Sun23PRBAndreev,Papaj23PRBAndreev,Nagae25PRBSpinpolarized} and proposed to function as a memory device~\cite{Giil24PRBSuperconductoraltermagnet}.  Altermagnetic Josephson junctions (AMJJs), such as superconductor/altermagnet/superconductor junctions, have been predicted to exhibit 0-$\pi$ transitions by modulating the junction length despite the vanishing net magnetization~\cite{Zhang24NCFinitemomentum,Ouassou23PRLDc,Beenakker23PRBPhaseshifted}. 
It has further been suggested that AMJJs support the formation of $\varphi$ junctions~\cite{Lu24PRLphi} and exhibit the superconducting diode effect~\cite{Banerjee24PRBAltermagnetic, Chakraborty24Perfect}. 

In this work, we study the effects of external electric or Zeeman fields on the Josephson response of AMJJs as shown in Fig.~\ref{Fig:setup}(a). 
We find that the current-phase relation (CPR) can be either forward or backward skewed due to a large second-order Josephson effect, which can be detected by double SQUID measurements~\cite{Messelot24PRLDirect}. The skewness changes with AFS parameter $t_J$ [Fig.~\ref{Fig:setup}(b)]. Near 0-$\pi$ transitions [plateau in Fig.~\ref{Fig:setup}(b)], supercurrent reversal emerges and the second-order Josephson effect dominates. When an electric field is applied to the altermagnetic region, for significant AFS, the critical current starts to oscillate around zero, leading to $0$-$\pi$ transitions. Interestingly, the magnitude of the critical current can be enhanced [Fig.~\ref{Fig:setup}(c)]. Moreover, external electric fields can also modify the CPR skewness [Fig.~\ref{Fig:setup}(c)]. Notably, the magnitude of the critical current can be substantially enhanced with increasing field strength for in-plane Zeeman fields [Fig.~\ref{Fig:setup}(d)]. Finally, we reveal that the Zeeman field can induce 0-$\pi$ transitions in AMJJs.

 {\color{blue}\emph{Model.}}---
%\section{Model}
Altermagnets feature spin splitting while having zero net magnetization, enforced by magnetic rotational or mirror symmetry, even in the absence of spin-orbit coupling. To be specific, we consider altermagnets with planar $d$-wave symmetry. It can be described by the tight-binding Hamiltonian on a square lattice~\cite{Smejkal22PRXGiant}
\begin{equation}
H({\bf k}) = t(\cos k_x + \cos k_y)\sigma_0 + t_J(\cos k_x -\cos k_y) \sigma_z,  
\label{Eq:AM_lattice}
\end{equation}
where ${\bf k} = (k_x,k_y)$ is the 2D momentum, $\sigma_{i}$ with $i\in\{x,y,z\}$ are Pauli matrices, and $\sigma_0$ is the identity matrices in spin space. The first term represents the normal kinetic energy, where $t=\hbar^2/2m$ with $m$ the effective electron mass in the continuum limit. The second term, parameterized by $t_J$, describes the altermagnetic order. Notably, the model obeys $\{C_2||C_{4z}\}$ spin symmetry, i.e., a combination of four-fold rotation in real space and two-fold rotation in spin space, 
which enforces $d$-wave planar magnetism in momentum space. 
There exist several candidate materials with planar $d$-wave altermagnetism, such as KRu$_4$O$_8$, La$_2$CuO$_4$ and FeSb$_2$~\cite{Smejkal22PRXConventional,Smejkal22PRXEmerging}. 
Transforming the Hamiltonian into real space yields
\begin{align}
\mathcal{H}_{\text{AM}} 
=
\sum_{\bf {r}} \left( \psi_{{\bf r}}^{\dagger} T_x \psi_{{\bf r}+ {\bm \delta}_x} + \psi_{{\bf r}}^{\dagger} T_y \psi_{{\bf r}+{\bm \delta}_y}  \psi_{{\bf r}} + H.c. \right),   
\label{eq:model-lattice}
\end{align}
where $ \psi_{\bf r} = (c_{{\bf r}\uparrow}, c_{{\bf r}\downarrow} )^T$ and $c_{{\bf r}\uparrow}\; (c_{{\bf r}\downarrow})$ is the electron annihilation operator at site $\bf r$ with spin $\uparrow ( \downarrow )$, ${\bm \delta}_x$ (${\bm \delta}_y$) is the unit vector in the $x$ ($y$) direction, and $T_x= t\sigma_0/2 + t_J \sigma_z/2 $ and $T_y=t\sigma_0/2 - t_J \sigma_z/2 $ are the hopping matrices in the $x$ and $y$ directions, respectively.

We consider planar Josephson junctions constructed by sandwiching the altermagnet with two conventional $s$-wave superconductors, as depicted in Fig.~\ref{Fig:setup}(a). The junction length and width are $L$ and $W$, respectively. This setup can be described by the following Bogoliubov-de-Gennes Hamiltonian
\begin{equation} \label{eq-ham-bdg}
{\cal H}_\text{BdG}({\bf r}) = 
\begin{pmatrix}
\mathcal{H} ({\bf r}) - \mu ({\bf r}) & i\Delta({\bf r}) \sigma_y \\
-i\Delta^\dagger({\bf r})\sigma_y & -\mathcal{H}^\ast({\bf r}) + \mu ({\bf r})
\end{pmatrix},
\end{equation}
where the position-dependent pairing potential is defined as $\Delta({\bf r}) = \Delta_0 e^{-i\;\text{sgn}(x)\phi/2}\Theta(|x|- L/2)$ with 
$\phi$ being the superconducting phase difference between the superconductors, $\Delta_0$ the pairing potential, $\Theta(x)$  the Heaviside function, and $\text{sgn}(x)$ the sign function. The chemical potential is $\mu({\bf r}) = \mu_{\text{AM}} \Theta(L/2-|x|)+\mu_{\text{S}} \Theta(|x|- L/2)$ with $\mu_{\text{AM}}$ and 
$\mu_{\text{S}}$ the chemical potentials in the altermagnetic and superconducting regions, respectively. The Hamiltonian $\mathcal{H}({\bf r})$ is given by
\begin{align}
\mathcal{H}({\bf r}) 
=
\mathcal{H}_{0}+\mathcal{H}_{\text{V}}+\mathcal{H}_{\text{Z}},
\end{align}
where $\mathcal{H}_{0}={\cal H}_{\text{AM}}$ in the altermagnetic region ($|x|<L/2$) while $\mathcal{H}_{0}=(t_{\bf{r} \rightarrow \bf{r} \pm \bf{\delta}_x} + t_{\bf{r} \rightarrow \bf{r} \pm \bf{\delta}_y} )t\sigma_0/2$ in the superconducting regions ($|x|\ge L/2$). $t_{\bf{r} \rightarrow \bf{r} \pm \bf{\delta}_x} $ and $t_{\bf{r} \rightarrow \bf{r} \pm \bf{\delta}_y}$ indicate hopping in the $x$ and $y$ directions, respectively. $\mathcal{H}_{\text{V}}=V_z \sigma_0 \Theta(L/2-|x|)$ is the electric potential with $V_z$ the magnitude and $H_Z=B_{y} \sigma_{y} \Theta(L/2-|x|)$ models the Zeeman field induced by the external magnetic field with $B_{y}$ the magnitude. Note that the electric potential and the magnetic field are applied in the middle altermagnetic region ($|x|<L/2$). Moreover, we consider an in-plane magnetic field so that the orbital effect of the magnetic field can be ignored.

To obtain the Josephson supercurrent, we adopt the well-established recursive Green's function approach~\cite{Furusaki94PBCMDC,Asano01PRBNumerical,ZhangSB2020PRB}, as detailed in the Appendix. For simplicity, periodic boundary conditions are applied along the $y$ direction, ensuring that $k_y$ remains a good quantum number. A dense $k_y$ grid (500 points) in the Brillouin zone is used in the calculation. For illustration and concreteness, we use the parameters $L=20$, $t=1$, $\mu_{\rm{S}}=\mu_{\rm{AM}}=0.2t$, $\Delta_0=0.01t$, and temperature $k_B T=0.02\Delta_0$ throughout the paper unless specified differently. 

\begin{figure}[!tp] 
\centering
\includegraphics[width=0.48\textwidth]{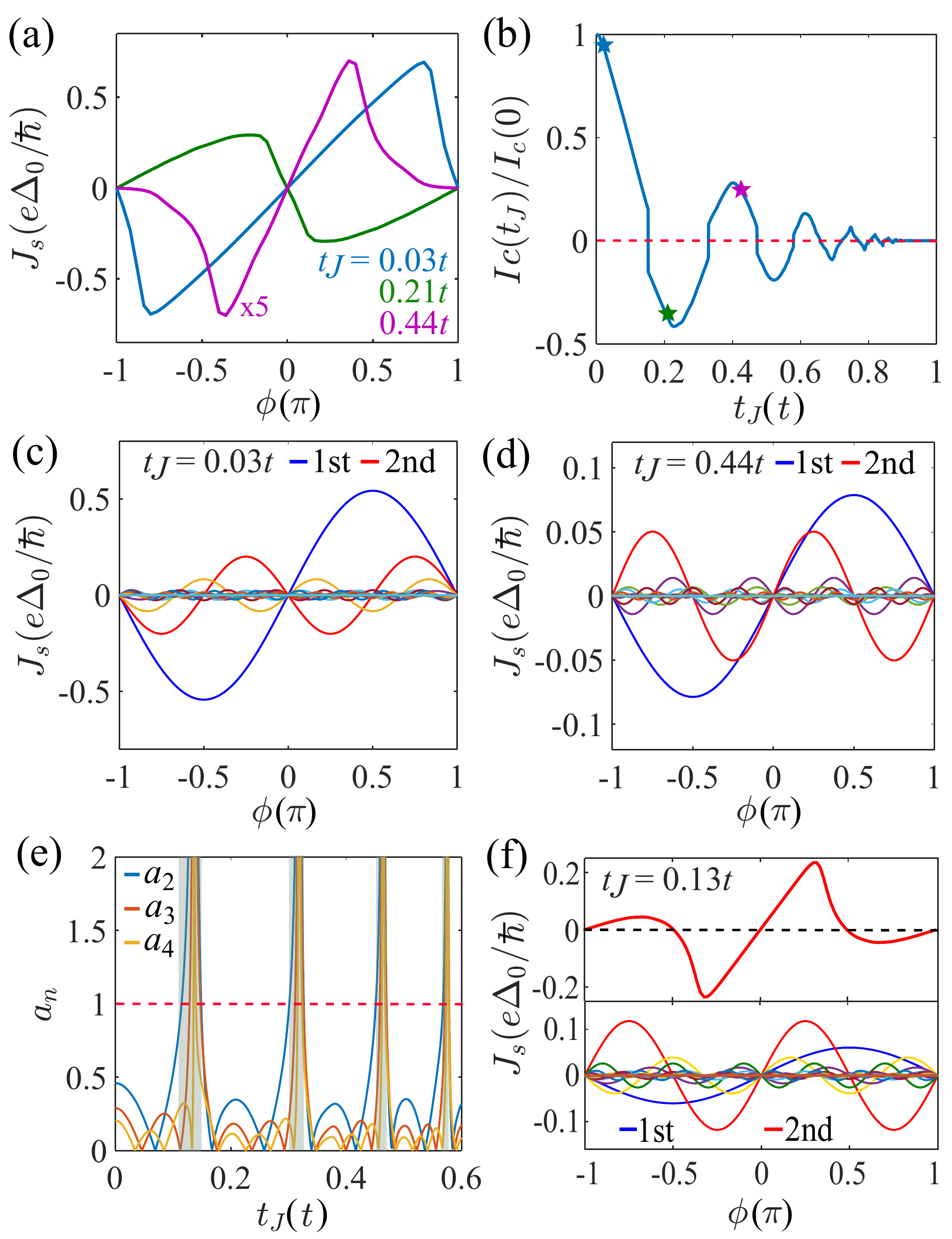}\\
\setlength{\abovecaptionskip}{-1pt}
\caption{(a) CPRs of the AMJJ for different $t_J$. Blue, green, and purple curves correspond to the AFS marked as stars in (b), respectively. 
(b) Critical current $I_c$ as a function of AFS $t_J$. 
(c) Harmonic expansion for $t_J$=0.03$t$. Dominant terms are first and second harmonics marked by blue and red colors, respectively. (d) Harmonic expansion for $t_J$=0.44$t$. Note the sign change of the second harmonic with respect to (c). (e) Higher harmonics $I_n$ compared to first harmonic $I_1$ as a function of AFS $t_J$ with $a_n \equiv I_n/I_1$. (f) Upper panel shows the CPR and lower panel illustrates dominant second harmonic for $t_J=0.13t$.
}
\label{Fig:AM_change}
\vspace{-1em}
\end{figure}

 {\color{blue}\emph{Altermagnetism induced tunable CPR skewness.}}---
%\section{Altermagnetism induced tunable CPR skewness}
We first analyze the Josephson current by changing AFS $t_J$. Figure~\ref{Fig:AM_change}(a) presents several typical CPRs for various $t_J$. All the CPRs are $2\pi$-periodic, i.e., $J_s(\phi)=J_s(\phi+2\pi)$, and odd in the phase difference $\phi$, i.e., $J_s(\phi)=J_s(-\phi)$, due to inversion symmetry of the junction. The supercurrent $J_s$ generally has either one maximum or minimum in the positive phase region $\phi\in(0, \pi)$, which we define as the critical current $I_c$. When the first harmonic dominates (e.g., in case of low-transparent junction interfaces), the supercurrent has a sinusoidal form $J_s=I_c \sin \phi$. Thus, the maximum or minimum is located at $\phi=\pi/2$. However, when the junction interfaces become more transparent, higher harmonics contribute, leading to skewed CPRs. The supercurrent is essentially determined by the derivative of the free energy $F$ of the junction to the phase difference $\phi$, $J_s\propto \partial F/\partial \phi$. For positive $I_c$, the ground state (corresponding to the free energy minimum) of the junction occurs at $\phi=0$. In contrast, for negative $I_c$, the ground state occurs at $\phi=\pi$. 

In Fig.~\ref{Fig:AM_change}(b), we calculate $I_c$ as a function of $t_J$.  We observe that $I_c$ oscillates around zero as $t_J$ increases. Similar results have been reported in a recent preprint exploiting complementary theoretical methods~\cite{Lu24PRLphi}. These oscillations indicate that the AMJJ undergoes 0-$\pi$ transitions [Fig.~\ref{Fig:AM_change}(a)]. They can be attributed to the proximity-induced finite-momentum pairing correlations in the altermagnet~\cite{Zhang24NCFinitemomentum}. The amplitude of $I_c$ first quickly decreases as the altermagnetism is turned on. Then, it oscillates with a reduced amplitude and finally vanishes for $t_J > 0.9t$. Note that $t_J$ is different from material to material and may be varied, e.g., by the twist angle in twisted bilayer platforms~\cite{He23PRLNonrelativistic,Sheoran24PRMNonrelativistic,Liu24PRLTwisted,Zeng24PRBBilayer}. 

Altermangetism can not only induce 0-$\pi$ transitions but also significantly affect the skewness of CPR where the impact of the second-order Josephson effect is vital. 
The CPR dominated by the first harmonic term has the form $J_s=I_c \sin \phi$. 
A skewed CPR implies the presence of pronounced higher harmonic contributing to the supercurrent, resulting in a non-sinusoidal form~\cite{Golubov04RMPCurrentphase}. The skewness can be characterized by the deviation of the maximum/minimum position $\phi_m$ of  $J_{s}$ from $\pi/2$, i.e., $\delta \phi \equiv \phi_m - \pi/2$. A positive $\delta \phi$ indicates a forward-skewed CPR, while a negative $\delta\phi$ corresponds to a backward-skewed CPR. 

In Fig.~\ref{Fig:setup}(b), we calculate the skewness of CPR as a function of AFS $t_{J}$. Strikingly, the CPR can be either forward (pink area) or backward skewed (green area). Examples of CPRs with forward and backward skewness are explicitly shown in Fig.~\ref{Fig:AM_change}(a). When the transparency of the junction changes because of the presence of altermagnetism, the second-order Josephson effect can not be ignored, leading to the non-sinusoidal form of Josephson supercurrent $J_s=I_{1} \sin(\phi) \pm I_{2} \sin(2\phi)$ with $I_{2}/I_1>0$. Note that the skewness of CPR is primarily dictated by the sign of the second harmonic term. Using 0 junction ($I_1>0$) as an example, for a negative second harmonic term, the CPR is forward-skewed with $J_s=I_{1} \sin(\phi) - I_{2} \sin(2\phi)$, as illustrated in Figs.~\ref{Fig:AM_change}(a) and \ref{Fig:AM_change}(c) with $t_J=0.03t$ and $I_2/I_1=0.37$. For a positive second harmonic term, the CPR is backward skewd with $J_s=I_{1} \sin(\phi) + I_{2} \sin(2\phi)$, as shown in Figs.~\ref{Fig:AM_change}(a) and \ref{Fig:AM_change}(d) with $t_J=0.44t$ and $I_2/I_1=0.64$. A similar ratio of $I_2/I_1$ has recently been observed in planar Josephson junctions based on Al/InAs-quantum wells~\cite{Zhang24SPLarge}. 
The second-order Josephson effect in AMJJs can be detected in SQUID devices or in tunnel junctions~\cite{Messelot24PRLDirect, Willsch24NPObservation}.
Notably, the skewness $\delta \phi$ can be as large as $\pm 0.3\pi$. More details on the skewness analysis can be found in the Appendx. 
This particular feature of tunable skewness of CPRs in AMJJs may be exploited to realize a large Josephson diode effect, e.g., in supercurrent interferometers~\cite{Souto22PRL}. 

We also note that multiple maxima may appear in the CPR due to dominant higher harmonics. Figure~\ref{Fig:AM_change}(e) shows the higher harmonic contribution compared to the first harmonic as a function of AFS $t_J$ with $I_n$ representing the amplitude of $n$th harmonic and $a_n \equiv I_n/I_1$. When higher harmonics are dominant ($a_n > 1$), as illustrated in the gray shaded area in Fig.~\ref{Fig:AM_change}(e), the AMJJ turns into a $\phi$-Josephson junction. An example for dominant second harmonic $a_2>1$ is shown in Fig.~\ref{Fig:AM_change}(f). In these cases, the skewness is not clearly defined. Thus, we exclude them in Fig.~\ref{Fig:setup}(b). We emphasize that the tunable second harmonic in the CPR is important in applications as it enhances the robustness of superconducting quantum bits~\cite{Messelot24PRLDirect}. Moreover, since the second harmonic can be accessed in experiments~\cite{Messelot24PRLDirect}, its sign change that leads to the skewness change can be exploited in superconducting circuits similar to controllable 0-$\pi$ junctions~\cite{Gingrich16NPControllable}. The altermagnet is robust against magnetic stray fields and has ultrafast spin dynamic~\cite{Smejkal22PRXEmerging}; thus the altermagnet-superconductor hybrid structure is advantageous for certain tasks in superconducting spintronics.

\begin{figure}[!tp] 
\centering
\includegraphics[width=0.49\textwidth]{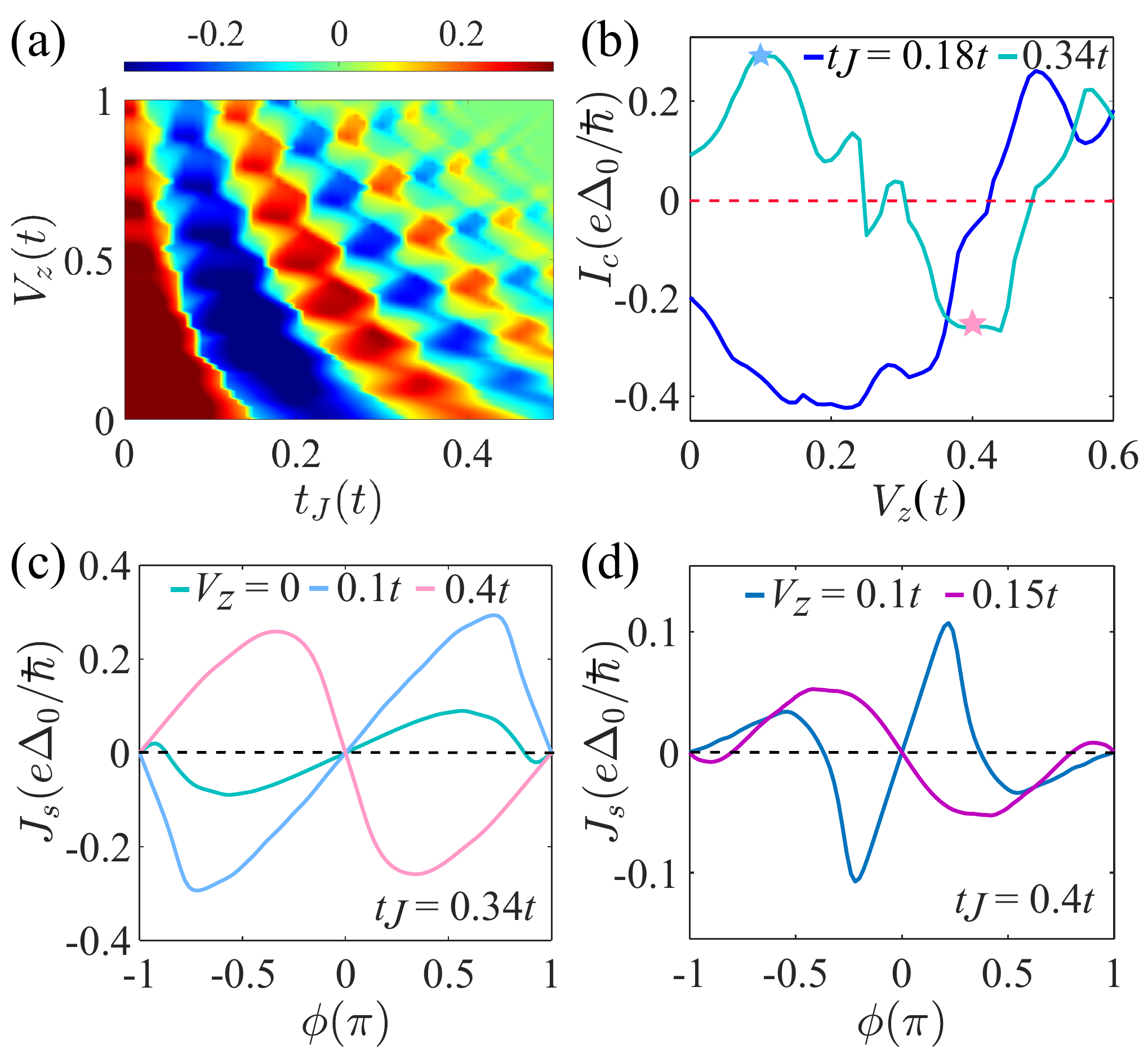}\\
\setlength{\abovecaptionskip}{-1pt}
\caption{(a) Phase diagram of the AMJJ as functions of AFS $t_J$ and applied gate potential $V_z$. The color bar stands for the magnitude of the critical current $I_c$ in units of $e\Delta_0/\hbar$. 
(b) $I_c$ as a function of $V_z$ for different $t_J$. The blue and cyan curves are for $t_J=0.18t$ and 0.34$t$, respectively. The light blue and pink stars indicate the $V_z$ values used in panel (c). (c) CPRs for different $V_z$ with $t_J=0.34t$. (d) CPRs of $\phi$-junctions for different $V_z$ with $t_J=0.4t$.
}
\label{Fig:V_change}
\vspace{-1em}
\end{figure}

 {\color{blue}\emph{Tuning CPR skewness and critical current by gating.}}---
%\section{Tuning CPR skewness and critical current by gating}
Now, we investigate the influence of an external gate potential $V_z$ on the Josephson current.
Figure~\ref{Fig:V_change}(a) shows the critical current $I_c$ of the AMJJ as functions of AFS $t_J$ and the gate potential $V_z$ for $\mu_{\text S}=\mu_{\text AM}$. Note that $V_z$ effectively tunes the chemical potential difference of superconductors and altermagnets.
For small $t_J$ (i.e., $t_J<0.05t$), $I_c$ is always positive with only slight oscillations. These oscillations appear due to Fabry-P\'erot interferences. Thus, there is no 0-$\pi$ transition. 

As $t_J$ increases, we observe $0$-$\pi$ transitions alongside a tilted interference pattern. The larger $t_J$, the more inclined the interference pattern becomes, as illustrated in the phase diagram in Fig.~\ref{Fig:V_change}(a). Hence, $I_c$ oscillates around zero with increasing $V_z$ for stronger AFS (i.e., $t_J>0.1t$). Figure~\ref{Fig:V_change}(b) presents $I_c$ as a function of $V_z$ for $t_J=0.18t$ and $0.34t$. This plot demonstrates the switching of the AMJJ between 0 and $\pi$ junctions by changing the gate potential, as illustrated in Fig.~\ref{Fig:V_change}(c). 
Meanwhile, the magnitude of $I_c$ is substantially enhanced for a wide range of gate potential, as shown in Fig.~\ref{Fig:V_change}(b). Notably, the critical current can be enhanced by a factor three before the 0-$\pi$ transition occurs for $V_z<0.2t$, as illustrated in Fig.~\ref{Fig:V_change}(b). The corresponding CPRs are displayed in Fig.~\ref{Fig:V_change}(c). The observed $I_c$ enhancement is primarily governed by finite-momentum pairing, depending on both the gate potential and the AFS~\cite{Zhang24NCFinitemomentum}.

Furthermore, the skewness of CPR can also be manipulated by electric gating. For instance, the CPR changes from forward to backward skewed, as shown in Fig.~\ref{Fig:setup}(c). Additionally, a 0 junction can be transformed into a $\pi$ or $\phi$ junction by adjusting the gate potential, as illustrated in Figs.~\ref{Fig:setup}(c) and \ref{Fig:V_change}(d). Gate-tunable $\phi$ junctions, shown in Fig.~\ref{Fig:V_change}(d), have previously been reported in Ref.~\cite{Lu24PRLphi}. These gate-tunable characteristics of AMJJs enrich their potential applications in superconducting circuits and spintronics. 

\begin{figure}[!tp] 
\centering
\includegraphics[width=0.49\textwidth]{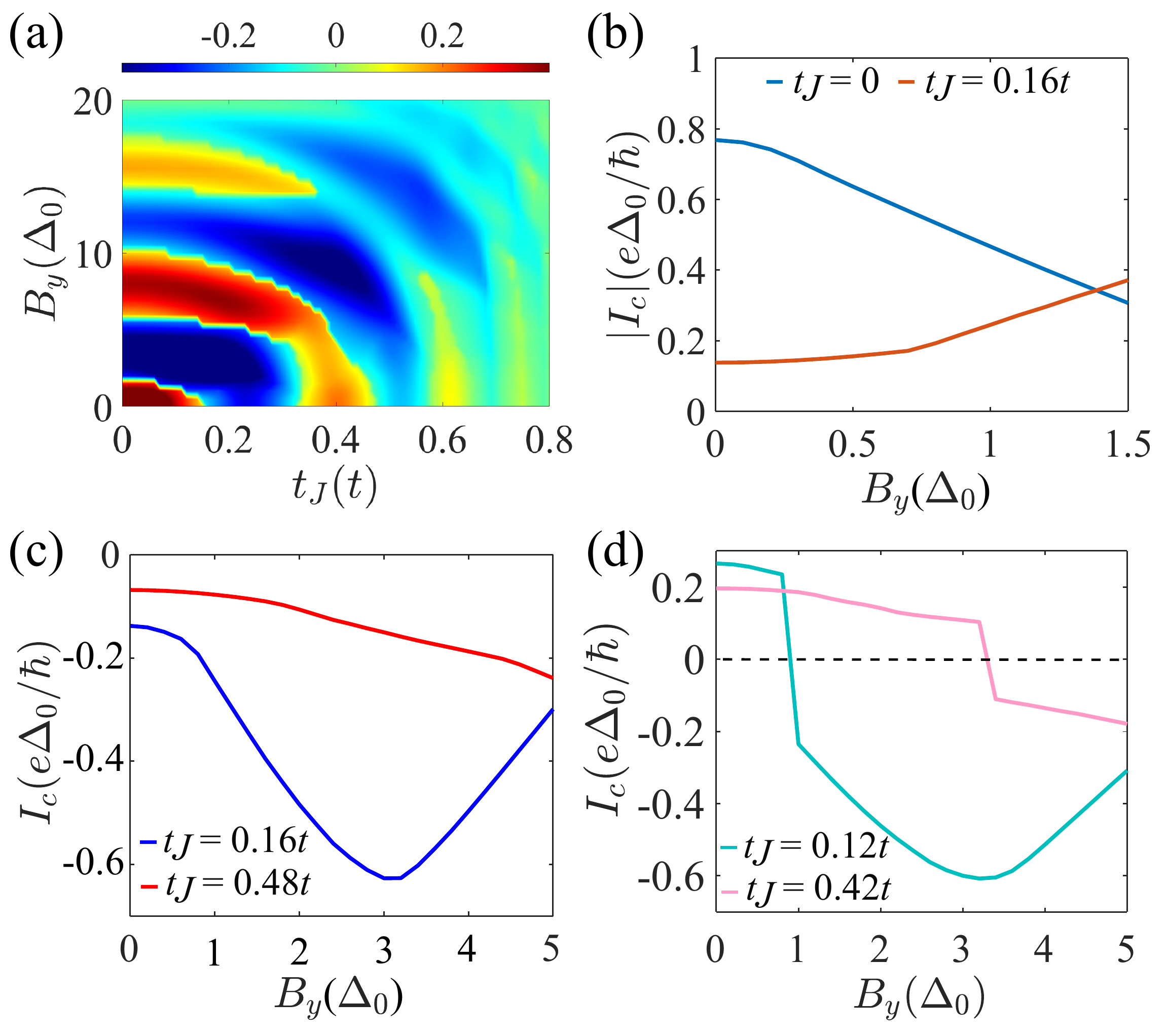}\\
\setlength{\abovecaptionskip}{-1pt}
\caption{(a) Phase diagram of the AMJJ as functions of $t_J$ (in units of $t$) and Zeeman field $B_y$ (in units of $\Delta_0$). The color bar indicates the magnitude of $I_c$ in units of $e\Delta_0/\hbar$. (b) $I_c$ (in units of $e\Delta_0/\hbar$) as a function of $B_y$ in absence and presence of altermagnetism. (c) $I_c$ as a function of $B_y$. The blue and red curves are for $t_J=0.16t$ and $0.48t$, respectively. (d) $I_c$ as a function of $B_y$. The cyan and pink curves are for $t_J=0.12t$ and $0.42t$, respectively. 
}
\label{Fig:B_change}
\vspace{-1em}
\end{figure}

 {\color{blue}\emph{Enhancement of supercurrent by Zeeman fields.}}---
%\section{Enhancement of supercurrent by Zeeman fields}
Now, we discuss the influence of an in-plane magnetic field on the supercurrent in AMJJs. The particular in-plane direction of the magnetic field does not matter qualitatively for our results.
Figure~\ref{Fig:B_change}(a) illustrates the phase diagram of the AMJJ against $t_J$ and $B_y$. The stripes bend downward as $t_J$ grows. This demonstrates the competition between altermagnetism and Zeeman effect. Strikingly, this competition gives rise to the enhancement of the critical current as shown in Figs.~\ref{Fig:setup}(d) and \ref{Fig:B_change}(b). In the absence of altermagnetism ($t_J=0$), the critical current $I_c$ decreases as the Zeeman field $B_y$ increases, illustrated by the blue line in Fig.\ref{Fig:B_change}(b). However, the critical current can be monotonically enhanced when both altermagnetism and Zeeman field are present, as illustrated in Fig.~\ref{Fig:setup}(d) and the orange line in Fig.~\ref{Fig:B_change}(b). Moreover, the magnitude of the critical current increases monotonically for large AFS as shown in Fig.~\ref{Fig:B_change}(c) for $t_J=0.48t$. Nevertheless, it first increases then decreases for small AFS, illustrated in Fig.~\ref{Fig:B_change}(c) for $t_J=0.16t$. To reach substantial ratios of $B_y/{\Delta_0}$, we envision that type-II superconductors, such as Nb and MoRe~\cite{Li21PRLTopological,Mandal24NPMagnetically}, form the AMJJs with an altermagnet with large g-factor as the weak link.

The enhancement of the critical current in the presence of an exchange field is uncommon because the critical current generally decreases with increasing field strength in both ferromagnetic~\cite{Ryazanov01PRLCoupling} and altermagnetic~[Fig.~\ref{Fig:AM_change}(b)] Josephson junctions. Nevertheless, in a superconductor/ferromagnet bilayer separated by an insulating film, the critical current can also be enhanced when the exchange fields in the two ferromagnets are antiparallel, resulting in no net magnetization~\cite{Bergeret01PRLEnhancement}. Notably, the mechanism underlying the enhancement of the critical current in AMJJs differs from this scenario, as AMJJs exhibit net magnetization in the presence of Zeeman fields. This phenomenon is a characteristic signature of altermagnetic superconducting heterostructures. 

Another impact of Zeeman fields on AMJJs is that it induces 0-$\pi$ transitions. $0$-$\pi$ transitions emerge as $B_y$ changes as shown in Fig.~\ref{Fig:B_change}(d). These oscillations stem from the Fermi surface spin splitting by the Zeeman field. For an AFS less than $0.2t$, the AMJJ transforms from a 0 junction ($I_c>0$) to a $\pi$ junction ($I_c<0$) for $B_y \leq \Delta_0$, as shown in Fig.~\ref{Fig:B_change}(a) and the cyan line in Fig.~\ref{Fig:B_change}(d). Additionally, it needs larger Zeeman field to induce 0-$\pi$ transition for large AFS as shown in Fig.~\ref{Fig:B_change}(d).

{\color{blue}\emph{Conclusion}.}---
% \section{Conclusion}
We investigate the influence of external electric and Zeeman fields on AMJJs. We reveal that changing the AFS significantly alters the second harmonic, enabling both forward- and backward-skewed CPRs in the junction. Applying either an electric or a Zeeman field can also significantly enhance the magnitude of the critical current across the junction. Remarkably, 0-$\pi$ transitions can be controlled by either electric gating or Zeeman fields. Moreover, we discover that the second harmonic in the CPR can be manipulated by electric gating.

\begin{acknowledgments} 
We thank Charles Gould, Lun-Hui Hu, Kristian Mæland, Martin Stehno and Shun Tamura for valuable discussions. This work was supported by EXC2147 ct.qmat (Project-ID: 390858490), the DFG-SFB 1170 (Project-ID: 258499086), and the Bavarian Ministry of Economic Affairs, Regional Development and Energy for ﬁnancial support within the High-Tech Agenda Project “Bausteine f\"ur das Quanten Computing auf Basis topologischer Materialen.” S.B.Z. acknowledges the start-up fund at HFNL, the Innovation Program for Quantum Science and Technology (Grant No. 2021ZD0302800), and Anhui Initiative in Quantum Information Technologies (Grant No. AHY170000).
\end{acknowledgments}

\appendix

\section{Lattice model}
We use the following continuum model of altermagnets~\cite{Smejkal22PRXGiant,Smejkal22PRXEmerging}:
\begin{eqnarray}
H(k) =  t(k_x^2+k_y^2)\sigma_0+t_J(k_y^2-k_x^2)\sigma_z.
\end{eqnarray}
By substituting $k_j^2 \rightarrow \frac{2}{a^2}[1 - \cos(k_j a)]$ with $j=x,y$ and $a=1$, we map the continuum model to a square lattice, which reads
\begin{eqnarray}
H( k_x, k_y) 
&= 
4t\sigma_0 - 2t[\cos(k_x)+\cos(k_y)]\sigma_0 \nonumber \\
&+ 2t_J [\cos(k_x)-\cos(k_y)]\sigma_z.
\end{eqnarray}
Ignoring the energy shift $4t$ and substituting $t^{\prime}=-2t$ and $t_J^{\prime}=2t_J$, we obtain the lattice model in the main text.
After Fourier transformation, we arrive at the real space Hamiltonian with periodic boundary conditions along the $y$ direction as
\begin{eqnarray}
\mathcal{H}\left( x, k_y \right) 
&=  
\sum_{x,x'} \{ \Psi_{x+a}^{\dagger} [ -t \sigma_0 + t_J \sigma_z ]  \Psi_{x'}  +  \Psi_{x}^{\dagger} [4t\sigma_0 \nonumber \\
&- 2t\cos k_y \sigma_0 - 2t_J \cos k_y \sigma_z ] \Psi_{x'} + \text{H.c.} \} \nonumber \\
\end{eqnarray}

\begin{figure*}[!htp] 
\centering
\includegraphics[width=0.90\textwidth]{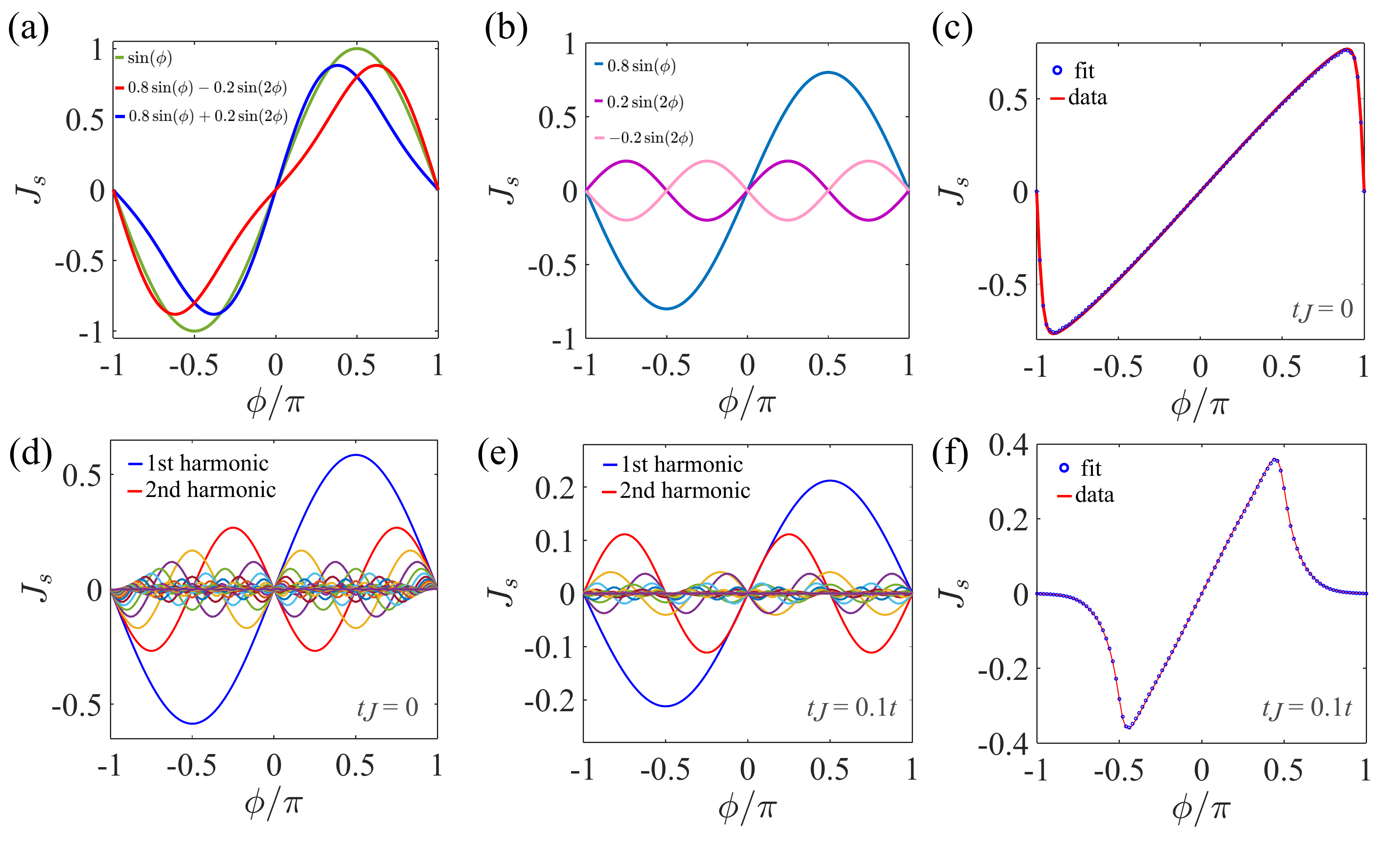}\\
\setlength{\abovecaptionskip}{-1pt}
\caption{(a) Josephson current $J_s$ as a function of superconducting phase difference $\phi$. Green curve indicates the sinusoidal case. Red and blue curves indicate the non-sinusoidal cases with negative and positive second harmonic terms $\sin(2\phi)$, respectively. (b) First and second harmonic terms, that contribute to the total Josephson current $J_s$, as a function of $\phi$. (c) CPR for $t_J=0$ with forward skewness. Red curve and blue circles represent the CPR data and harmonic fit, respectively.  (d) Harmonic components for $t_J$=0. Dominant terms are first and second harmonic contributions marked by blue and red colors, respectively. (e) The harmonic components for the case with $t_J$=0.1$t$. The dominant terms are first and second harmonic contributions marked by blue and red colors, respectively. (f) CPR for $t_J=0.1t$ with backward skewness. Red curve and blue circles represent the CPR data and harmonic fit, respectively.
}
\label{Fig:SM_harmonic}
\vspace{-1em}
\end{figure*}

\section{Josephson current}
The Josephson current can be obtained by the tunnel Hamiltonian~\cite{Cohen62PRLSuperconductive,Ambegaokar63PRLTunneling}. In the middle region, charge conservation still holds. Thus, the total current $J_s$ across the junction is defined as the rate of change of the number of electrons (${\partial N}/{\partial t}$) in the middle region times the electron charge~\cite{Abrikosov12Methods}: 
\begin{eqnarray}
J_s 
&=& 
-e \frac{\partial N}{\partial t}  \nonumber \\
&=&
\frac{ie}{\hbar}  \sum_{\sigma} ( t_{j,j+1} \langle c_{j,\sigma}^{\dagger} c_{j+1,\sigma} \rangle 
- t_{j+1,j} \langle c_{j+1,\sigma}^{\dagger} c_{j,\sigma} \rangle ) \label{Eq:J_s^e} 
 \nonumber \\
\end{eqnarray} 
The above equation can be rewritten in terms of Green's function as 
\begin{align}
J_s 
&=
\frac{i e T}{\hbar} \sum_{\omega_n} \sum_{k=0}^{n_y} {\rm Tr} [\tau_z T_{j,j+1} G_{\omega_n}(j,k;j+1,k) \nonumber \\
&- \tau_z T_{j+1,j} G_{\omega_n}(j+1,k;j,k)],
\label{Eq:Js}
\end{align}
where $G_{\omega_n}(j,k;j^{\prime},k^{\prime})$ is the Green's function in Nambu space, $\tau_z$ is the third Pauli matrix in particle-hole space, $\omega_n=(2n+1) \pi T $ is the Matsubara frequency, and $T_{j,j^{\prime}}$ is the hopping matrix. We calculate the Josephson current exploiting the recursive Green’s function approach~\cite{Furusaki94PBCMDC,Asano01PRBNumerical}. Since we assume translation symmetry along the $y$ direction, $k_y$ is a good quantum number. The planar Josephson junction we study is a quasi-1D system along the $x$ direction. In the calculation, we put the Boltzmann constant $k_B=1$. The number of $k_y$ points is 500.
The Nambu Green's function for each Matsubara frequency and momentum $k_y$ can be obtained by the iterative method
 \begin{align}
 &G_{\omega_n}(j,k;j^{\prime},k^{\prime}) 
 =[i\omega_n \delta_{k,k^{\prime}} I - H_{0} \nonumber \\
 &- H_{1} G_{\omega_n} (j-1,k; j-1, k^{\prime}) H_{1} ]^{-1},
 \end{align}
%\begin{align}
%G_{\omega_n}(j,k;j^{\prime},k^{\prime}) 
%&=[i\omega_n \delta_{k,k^{\prime}} I - H_{0} 
%- H_{1} \nonumber \\
%& \times G_{\omega_n} (j-1,k; j-1, k^{\prime}) H_{1} ]^{-1},
%\end{align}
where $I$ is a $4\times4$ unit matrix, $H_{0}$ is the on-site energy matrix, and $H_1$ is the hopping matrix between two neighboring sites.
The superconducting leads are modeled using self-energy terms attached to the altermagnetic region. The self-energy terms incorporate the superconducting gap $\Delta_0$ and the phase difference $\phi$ between the superconductors. By attaching the superconductor to the altermagnetic region, the Green's function of the connected system can be obtained as 
\begin{align}
&G_{\omega_n}(j,k;j+1,k^{\prime}) 
=\sum_{l=0}^{n_y} \{G_{\omega_n}(j,k;j,l)^{-1} 
\nonumber \\
&-H_1 G_{\omega_n}(j+1,k;j+1,l) H_1 \}^{-1} H_{1} G_{\omega_n}(j+1,l;j+1,k^{\prime}), \\
&G_{\omega_n}(j+1,k;j,k^{\prime}) 
=\sum_{l=0}^{n_y} G_{\omega_n}(j+1,k;j+1,l) H_{1}  \nonumber \\ 
&\times \{G_{\omega_n}(j,l;j,k^{\prime})^{-1} 
- H_1 G_{\omega_n}(j+1,l;j+1,k^{\prime}) H_1 \}^{-1}.
\end{align}
By substituting the above two equations into Eq.(\ref{Eq:Js}) for each Matsubara frequency and then summing over them, we obtain the dc Josephson current. For each phase difference $\phi$, we obtain the Josephson current $J_s(\phi)$. Plotting the Josephson current $J_s(\phi)$ as a function of $\phi$, we arrive at the current phase relation (CPR) as shown in Fig.\ref{Fig:setup}(c). For every CPR, it can be expanded into Fourier series to obtain the contribution of each harmonic as shown in Fig.~\ref{Fig:AM_change}(e) and detailed in the next section. 

\section{Skewness analysis}
In the main text, we show that the skewness varies as the altermagnetic field strength grows (Fig.1 in the main text). Here, we propose a simple model to simulate the skewness change. The sinusoidal form of the Josephson supercurrent reads $J_s=I_c \sin(\phi)$ with $I_c$ the critical current, shown in Fig.~\ref{Fig:SM_harmonic}(a). When the transparency of the junction changes because of the introduction of altermagnetism, the second harmonic contribution can not be ignored leading to the nonsinusoidal form of Josephson supercurrent $J_s=I_{1} \sin(\phi) \pm I_{2} \sin(2\phi)$ with $I_{2}>0$. For the case with negative second harmonic term, the skewness is forward with $J_s=I_{1} \sin(\phi) - I_{2} \sin(2\phi)$ as illustrated in Fig.~\ref{Fig:SM_harmonic}(a). For the case with positive second harmonic term, the skewness is backward with $J_s=I_{1} \sin(\phi) + I_{2} \sin(2\phi)$. The first harmonic and second harmonic terms are depicted in Fig.~\ref{Fig:SM_harmonic}(b). 
The nonsinusoidal behavior without altermagnetism ($t_J=0$) can be captured by expressing the CPR as Fourier series~\cite{Golubov04RMPCurrentphase}
\begin{eqnarray}
J_s(\phi) 
&=& \sum_n (-1)^{n-1} I_n \sin(n \phi) \nonumber \\
&=& \sum_n (-1)^{n-1} I_1 a_n \sin(n \phi)
\end{eqnarray}
where $n$ is an integer, $a_n \equiv I_n/I_1$ and $I_n>0$.
It is illustrated in Figs.~\ref{Fig:SM_harmonic} (c) and (d) with $a_2=I_2/I_1=0.46$. When the altermagnetism is introduced, the higher harmonic terms change sign. For instance, as the sign of the second harmonic term ($n=2$) changes from minus to plus the skewness of the CPR becomes backward as illustrated in Figs.~\ref{Fig:SM_harmonic} (e) and (f) with $a_2=I_2/I_1=0.53$.


%\bibliographystyle{apsrev4-2}
%\bibliography{Library-Sun}

\begin{thebibliography}{60}%
\makeatletter
\providecommand \@ifxundefined [1]{%
 \@ifx{#1\undefined}
}%
\providecommand \@ifnum [1]{%
 \ifnum #1\expandafter \@firstoftwo
 \else \expandafter \@secondoftwo
 \fi
}%
\providecommand \@ifx [1]{%
 \ifx #1\expandafter \@firstoftwo
 \else \expandafter \@secondoftwo
 \fi
}%
\providecommand \natexlab [1]{#1}%
\providecommand \enquote  [1]{``#1''}%
\providecommand \bibnamefont  [1]{#1}%
\providecommand \bibfnamefont [1]{#1}%
\providecommand \citenamefont [1]{#1}%
\providecommand \href@noop [0]{\@secondoftwo}%
\providecommand \href [0]{\begingroup \@sanitize@url \@href}%
\providecommand \@href[1]{\@@startlink{#1}\@@href}%
\providecommand \@@href[1]{\endgroup#1\@@endlink}%
\providecommand \@sanitize@url [0]{\catcode `\\12\catcode `\$12\catcode
  `\&12\catcode `\#12\catcode `\^12\catcode `\_12\catcode `\%12\relax}%
\providecommand \@@startlink[1]{}%
\providecommand \@@endlink[0]{}%
\providecommand \url  [0]{\begingroup\@sanitize@url \@url }%
\providecommand \@url [1]{\endgroup\@href {#1}{\urlprefix }}%
\providecommand \urlprefix  [0]{URL }%
\providecommand \Eprint [0]{\href }%
\providecommand \doibase [0]{https://doi.org/}%
\providecommand \selectlanguage [0]{\@gobble}%
\providecommand \bibinfo  [0]{\@secondoftwo}%
\providecommand \bibfield  [0]{\@secondoftwo}%
\providecommand \translation [1]{[#1]}%
\providecommand \BibitemOpen [0]{}%
\providecommand \bibitemStop [0]{}%
\providecommand \bibitemNoStop [0]{.\EOS\space}%
\providecommand \EOS [0]{\spacefactor3000\relax}%
\providecommand \BibitemShut  [1]{\csname bibitem#1\endcsname}%
\let\auto@bib@innerbib\@empty
%</preamble>
\bibitem [{\citenamefont {{\v S}mejkal}\ \emph
  {et~al.}(2022{\natexlab{a}})\citenamefont {{\v S}mejkal}, \citenamefont
  {Sinova},\ and\ \citenamefont {Jungwirth}}]{Smejkal22PRXEmerging}%
  \BibitemOpen
  \bibfield  {author} {\bibinfo {author} {\bibfnamefont {L.}~\bibnamefont {{\v
  S}mejkal}}, \bibinfo {author} {\bibfnamefont {J.}~\bibnamefont {Sinova}},\
  and\ \bibinfo {author} {\bibfnamefont {T.}~\bibnamefont {Jungwirth}},\ }\href
  {https://doi.org/10.1103/PhysRevX.12.040501} {\bibfield  {journal} {\bibinfo
  {journal} {Phys. Rev. X}\ }\textbf {\bibinfo {volume} {12}},\ \bibinfo
  {pages} {040501} (\bibinfo {year} {2022}{\natexlab{a}})}\BibitemShut
  {NoStop}%
\bibitem [{\citenamefont {{\v S}mejkal}\ \emph
  {et~al.}(2022{\natexlab{b}})\citenamefont {{\v S}mejkal}, \citenamefont
  {Sinova},\ and\ \citenamefont {Jungwirth}}]{Smejkal22PRXConventional}%
  \BibitemOpen
  \bibfield  {author} {\bibinfo {author} {\bibfnamefont {L.}~\bibnamefont {{\v
  S}mejkal}}, \bibinfo {author} {\bibfnamefont {J.}~\bibnamefont {Sinova}},\
  and\ \bibinfo {author} {\bibfnamefont {T.}~\bibnamefont {Jungwirth}},\ }\href
  {https://doi.org/10.1103/PhysRevX.12.031042} {\bibfield  {journal} {\bibinfo
  {journal} {Phys. Rev. X}\ }\textbf {\bibinfo {volume} {12}},\ \bibinfo
  {pages} {031042} (\bibinfo {year} {2022}{\natexlab{b}})}\BibitemShut
  {NoStop}%
\bibitem [{\citenamefont {Wu}\ \emph {et~al.}(2007)\citenamefont {Wu},
  \citenamefont {Sun}, \citenamefont {Fradkin},\ and\ \citenamefont
  {Zhang}}]{CJWu07PRB}%
  \BibitemOpen
  \bibfield  {author} {\bibinfo {author} {\bibfnamefont {C.}~\bibnamefont
  {Wu}}, \bibinfo {author} {\bibfnamefont {K.}~\bibnamefont {Sun}}, \bibinfo
  {author} {\bibfnamefont {E.}~\bibnamefont {Fradkin}},\ and\ \bibinfo {author}
  {\bibfnamefont {S.-C.}\ \bibnamefont {Zhang}},\ }\href
  {https://doi.org/10.1103/PhysRevB.75.115103} {\bibfield  {journal} {\bibinfo
  {journal} {Phys. Rev. B}\ }\textbf {\bibinfo {volume} {75}},\ \bibinfo
  {pages} {115103} (\bibinfo {year} {2007})}\BibitemShut {NoStop}%
\bibitem [{\citenamefont {Naka}\ \emph {et~al.}(2019)\citenamefont {Naka},
  \citenamefont {Hayami}, \citenamefont {Kusunose}, \citenamefont {Yanagi},
  \citenamefont {Motome},\ and\ \citenamefont {Seo}}]{Naka19NCSpin}%
  \BibitemOpen
  \bibfield  {author} {\bibinfo {author} {\bibfnamefont {M.}~\bibnamefont
  {Naka}}, \bibinfo {author} {\bibfnamefont {S.}~\bibnamefont {Hayami}},
  \bibinfo {author} {\bibfnamefont {H.}~\bibnamefont {Kusunose}}, \bibinfo
  {author} {\bibfnamefont {Y.}~\bibnamefont {Yanagi}}, \bibinfo {author}
  {\bibfnamefont {Y.}~\bibnamefont {Motome}},\ and\ \bibinfo {author}
  {\bibfnamefont {H.}~\bibnamefont {Seo}},\ }\href
  {https://doi.org/10.1038/s41467-019-12229-y} {\bibfield  {journal} {\bibinfo
  {journal} {Nat. Commun.}\ }\textbf {\bibinfo {volume} {10}},\ \bibinfo
  {pages} {4305} (\bibinfo {year} {2019})}\BibitemShut {NoStop}%
\bibitem [{\citenamefont {Shao}\ \emph {et~al.}(2021)\citenamefont {Shao},
  \citenamefont {Zhang}, \citenamefont {Li}, \citenamefont {Eom},\ and\
  \citenamefont {Tsymbal}}]{shao2021spin}%
  \BibitemOpen
  \bibfield  {author} {\bibinfo {author} {\bibfnamefont {D.-F.}\ \bibnamefont
  {Shao}}, \bibinfo {author} {\bibfnamefont {S.-H.}\ \bibnamefont {Zhang}},
  \bibinfo {author} {\bibfnamefont {M.}~\bibnamefont {Li}}, \bibinfo {author}
  {\bibfnamefont {C.-B.}\ \bibnamefont {Eom}},\ and\ \bibinfo {author}
  {\bibfnamefont {E.~Y.}\ \bibnamefont {Tsymbal}},\ }\href
  {https://doi.org/10.1038/s41467-021-26915-3} {\bibfield  {journal} {\bibinfo
  {journal} {Nat. Commun.}\ }\textbf {\bibinfo {volume} {12}},\ \bibinfo
  {pages} {7061} (\bibinfo {year} {2021})}\BibitemShut {NoStop}%
\bibitem [{\citenamefont {Hayami}\ \emph {et~al.}(2019)\citenamefont {Hayami},
  \citenamefont {Yanagi},\ and\ \citenamefont
  {Kusunose}}]{Hayami19JPSJMomentumDependent}%
  \BibitemOpen
  \bibfield  {author} {\bibinfo {author} {\bibfnamefont {S.}~\bibnamefont
  {Hayami}}, \bibinfo {author} {\bibfnamefont {Y.}~\bibnamefont {Yanagi}},\
  and\ \bibinfo {author} {\bibfnamefont {H.}~\bibnamefont {Kusunose}},\ }\href
  {https://doi.org/10.7566/JPSJ.88.123702} {\bibfield  {journal} {\bibinfo
  {journal} {J. Phys. Soc. Jpn.}\ }\textbf {\bibinfo {volume} {88}},\ \bibinfo
  {pages} {123702} (\bibinfo {year} {2019})}\BibitemShut {NoStop}%
\bibitem [{\citenamefont {Ahn}\ \emph {et~al.}(2019)\citenamefont {Ahn},
  \citenamefont {Hariki}, \citenamefont {Lee},\ and\ \citenamefont {Kune{\v
  s}}}]{Ahn19PRBAntiferromagnetism}%
  \BibitemOpen
  \bibfield  {author} {\bibinfo {author} {\bibfnamefont {K.-H.}\ \bibnamefont
  {Ahn}}, \bibinfo {author} {\bibfnamefont {A.}~\bibnamefont {Hariki}},
  \bibinfo {author} {\bibfnamefont {K.-W.}\ \bibnamefont {Lee}},\ and\ \bibinfo
  {author} {\bibfnamefont {J.}~\bibnamefont {Kune{\v s}}},\ }\href
  {https://doi.org/10.1103/PhysRevB.99.184432} {\bibfield  {journal} {\bibinfo
  {journal} {Phys. Rev. B}\ }\textbf {\bibinfo {volume} {99}},\ \bibinfo
  {pages} {184432} (\bibinfo {year} {2019})}\BibitemShut {NoStop}%
\bibitem [{\citenamefont {{\v S}mejkal}\ \emph {et~al.}(2020)\citenamefont {{\v
  S}mejkal}, \citenamefont {{Gonz{\'a}lez-Hern{\'a}ndez}}, \citenamefont
  {Jungwirth},\ and\ \citenamefont {Sinova}}]{Smejkal20SACrystal}%
  \BibitemOpen
  \bibfield  {author} {\bibinfo {author} {\bibfnamefont {L.}~\bibnamefont {{\v
  S}mejkal}}, \bibinfo {author} {\bibfnamefont {R.}~\bibnamefont
  {{Gonz{\'a}lez-Hern{\'a}ndez}}}, \bibinfo {author} {\bibfnamefont
  {T.}~\bibnamefont {Jungwirth}},\ and\ \bibinfo {author} {\bibfnamefont
  {J.}~\bibnamefont {Sinova}},\ }\href {https://doi.org/10.1126/sciadv.aaz8809}
  {\bibfield  {journal} {\bibinfo  {journal} {Sci. Adv.}\ }\textbf {\bibinfo
  {volume} {6}},\ \bibinfo {pages} {eaaz8809} (\bibinfo {year}
  {2020})}\BibitemShut {NoStop}%
\bibitem [{\citenamefont {Yuan}\ \emph {et~al.}(2020)\citenamefont {Yuan},
  \citenamefont {Wang}, \citenamefont {Luo}, \citenamefont {Rashba},\ and\
  \citenamefont {Zunger}}]{Yuan20PRBGiant}%
  \BibitemOpen
  \bibfield  {author} {\bibinfo {author} {\bibfnamefont {L.-D.}\ \bibnamefont
  {Yuan}}, \bibinfo {author} {\bibfnamefont {Z.}~\bibnamefont {Wang}}, \bibinfo
  {author} {\bibfnamefont {J.-W.}\ \bibnamefont {Luo}}, \bibinfo {author}
  {\bibfnamefont {E.~I.}\ \bibnamefont {Rashba}},\ and\ \bibinfo {author}
  {\bibfnamefont {A.}~\bibnamefont {Zunger}},\ }\href
  {https://doi.org/10.1103/PhysRevB.102.014422} {\bibfield  {journal} {\bibinfo
   {journal} {Phys. Rev. B}\ }\textbf {\bibinfo {volume} {102}},\ \bibinfo
  {pages} {014422} (\bibinfo {year} {2020})}\BibitemShut {NoStop}%
\bibitem [{\citenamefont {Bai}\ \emph {et~al.}(2024)\citenamefont {Bai},
  \citenamefont {Feng}, \citenamefont {Liu}, \citenamefont {{\v S}mejkal},
  \citenamefont {Mokrousov},\ and\ \citenamefont {Yao}}]{Bai24Altermagnetism}%
  \BibitemOpen
  \bibfield  {author} {\bibinfo {author} {\bibfnamefont {L.}~\bibnamefont
  {Bai}}, \bibinfo {author} {\bibfnamefont {W.}~\bibnamefont {Feng}}, \bibinfo
  {author} {\bibfnamefont {S.}~\bibnamefont {Liu}}, \bibinfo {author}
  {\bibfnamefont {L.}~\bibnamefont {{\v S}mejkal}}, \bibinfo {author}
  {\bibfnamefont {Y.}~\bibnamefont {Mokrousov}},\ and\ \bibinfo {author}
  {\bibfnamefont {Y.}~\bibnamefont {Yao}},\ }\href
  {https://doi.org/10.1002/adfm.202409327} {\bibfield  {journal} {\bibinfo
  {journal} {Adv. Funct. Mater.}\ }\textbf {\bibinfo {volume} {34}},\ \bibinfo
  {pages} {2409327} (\bibinfo {year} {2024})}\BibitemShut {NoStop}%
\bibitem [{\citenamefont {Berlijn}\ \emph {et~al.}(2017)\citenamefont
  {Berlijn}, \citenamefont {Snijders}, \citenamefont {Delaire}, \citenamefont
  {Zhou}, \citenamefont {Maier}, \citenamefont {Cao}, \citenamefont {Chi},
  \citenamefont {Matsuda}, \citenamefont {Wang}, \citenamefont {Koehler},
  \citenamefont {Kent},\ and\ \citenamefont
  {Weitering}}]{Berlijn17PRLItinerant}%
  \BibitemOpen
  \bibfield  {author} {\bibinfo {author} {\bibfnamefont {T.}~\bibnamefont
  {Berlijn}}, \bibinfo {author} {\bibfnamefont {P.~C.}\ \bibnamefont
  {Snijders}}, \bibinfo {author} {\bibfnamefont {O.}~\bibnamefont {Delaire}},
  \bibinfo {author} {\bibfnamefont {H.-D.}\ \bibnamefont {Zhou}}, \bibinfo
  {author} {\bibfnamefont {T.~A.}\ \bibnamefont {Maier}}, \bibinfo {author}
  {\bibfnamefont {H.-B.}\ \bibnamefont {Cao}}, \bibinfo {author} {\bibfnamefont
  {S.-X.}\ \bibnamefont {Chi}}, \bibinfo {author} {\bibfnamefont
  {M.}~\bibnamefont {Matsuda}}, \bibinfo {author} {\bibfnamefont
  {Y.}~\bibnamefont {Wang}}, \bibinfo {author} {\bibfnamefont {M.~R.}\
  \bibnamefont {Koehler}}, \bibinfo {author} {\bibfnamefont {P.~R.~C.}\
  \bibnamefont {Kent}},\ and\ \bibinfo {author} {\bibfnamefont {H.~H.}\
  \bibnamefont {Weitering}},\ }\href
  {https://doi.org/10.1103/PhysRevLett.118.077201} {\bibfield  {journal}
  {\bibinfo  {journal} {Phys. Rev. Lett.}\ }\textbf {\bibinfo {volume} {118}},\
  \bibinfo {pages} {077201} (\bibinfo {year} {2017})}\BibitemShut {NoStop}%
\bibitem [{\citenamefont {Mazin}\ \emph {et~al.}(2021)\citenamefont {Mazin},
  \citenamefont {Koepernik}, \citenamefont {Johannes}, \citenamefont
  {{Gonz{\'a}lez-Hern{\'a}ndez}},\ and\ \citenamefont {{\v
  S}mejkal}}]{Mazin21PNASPrediction}%
  \BibitemOpen
  \bibfield  {author} {\bibinfo {author} {\bibfnamefont {I.~I.}\ \bibnamefont
  {Mazin}}, \bibinfo {author} {\bibfnamefont {K.}~\bibnamefont {Koepernik}},
  \bibinfo {author} {\bibfnamefont {M.~D.}\ \bibnamefont {Johannes}}, \bibinfo
  {author} {\bibfnamefont {R.}~\bibnamefont {{Gonz{\'a}lez-Hern{\'a}ndez}}},\
  and\ \bibinfo {author} {\bibfnamefont {L.}~\bibnamefont {{\v S}mejkal}},\
  }\href {https://doi.org/10.1073/pnas.2108924118} {\bibfield  {journal}
  {\bibinfo  {journal} {Proc. Natl. Acad. Sci. U.S.A.}\ }\textbf {\bibinfo
  {volume} {118}},\ \bibinfo {pages} {e2108924118} (\bibinfo {year}
  {2021})}\BibitemShut {NoStop}%
\bibitem [{\citenamefont {{\v S}mejkal}\ \emph
  {et~al.}(2022{\natexlab{c}})\citenamefont {{\v S}mejkal}, \citenamefont
  {Hellenes}, \citenamefont {{Gonz{\'a}lez-Hern{\'a}ndez}}, \citenamefont
  {Sinova},\ and\ \citenamefont {Jungwirth}}]{Smejkal22PRXGiant}%
  \BibitemOpen
  \bibfield  {author} {\bibinfo {author} {\bibfnamefont {L.}~\bibnamefont {{\v
  S}mejkal}}, \bibinfo {author} {\bibfnamefont {A.~B.}\ \bibnamefont
  {Hellenes}}, \bibinfo {author} {\bibfnamefont {R.}~\bibnamefont
  {{Gonz{\'a}lez-Hern{\'a}ndez}}}, \bibinfo {author} {\bibfnamefont
  {J.}~\bibnamefont {Sinova}},\ and\ \bibinfo {author} {\bibfnamefont
  {T.}~\bibnamefont {Jungwirth}},\ }\href
  {https://doi.org/10.1103/PhysRevX.12.011028} {\bibfield  {journal} {\bibinfo
  {journal} {Phys. Rev. X}\ }\textbf {\bibinfo {volume} {12}},\ \bibinfo
  {pages} {011028} (\bibinfo {year} {2022}{\natexlab{c}})}\BibitemShut
  {NoStop}%
\bibitem [{\citenamefont {Mazin}(2023)}]{Mazin23PRBAltermagnetism}%
  \BibitemOpen
  \bibfield  {author} {\bibinfo {author} {\bibfnamefont {I.~I.}\ \bibnamefont
  {Mazin}},\ }\href {https://doi.org/10.1103/PhysRevB.107.L100418} {\bibfield
  {journal} {\bibinfo  {journal} {Phys. Rev. B}\ }\textbf {\bibinfo {volume}
  {107}},\ \bibinfo {pages} {L100418} (\bibinfo {year} {2023})}\BibitemShut
  {NoStop}%
\bibitem [{\citenamefont {Guo}\ \emph {et~al.}(2023)\citenamefont {Guo},
  \citenamefont {Liu}, \citenamefont {Janson}, \citenamefont {Fulga},
  \citenamefont {{van den Brink}},\ and\ \citenamefont
  {Facio}}]{Guo23MTPSpinsplit}%
  \BibitemOpen
  \bibfield  {author} {\bibinfo {author} {\bibfnamefont {Y.}~\bibnamefont
  {Guo}}, \bibinfo {author} {\bibfnamefont {H.}~\bibnamefont {Liu}}, \bibinfo
  {author} {\bibfnamefont {O.}~\bibnamefont {Janson}}, \bibinfo {author}
  {\bibfnamefont {I.~C.}\ \bibnamefont {Fulga}}, \bibinfo {author}
  {\bibfnamefont {J.}~\bibnamefont {{van den Brink}}},\ and\ \bibinfo {author}
  {\bibfnamefont {J.~I.}\ \bibnamefont {Facio}},\ }\href
  {https://doi.org/10.1016/j.mtphys.2023.100991} {\bibfield  {journal}
  {\bibinfo  {journal} {Mater. Today Phys.}\ }\textbf {\bibinfo {volume}
  {32}},\ \bibinfo {pages} {100991} (\bibinfo {year} {2023})}\BibitemShut
  {NoStop}%
\bibitem [{\citenamefont {{Lin}}\ \emph {et~al.}()\citenamefont {{Lin}},
  \citenamefont {{Chen}}, \citenamefont {{Lu}}, \citenamefont {{Liang}},
  \citenamefont {{Feng}}, \citenamefont {{Yamagami}}, \citenamefont
  {{Osiecki}}, \citenamefont {{Leandersson}}, \citenamefont {{Thiagarajan}},
  \citenamefont {{Liu}}, \citenamefont {{Felser}},\ and\ \citenamefont
  {{Ma}}}]{Lin24Observation}%
  \BibitemOpen
  \bibfield  {author} {\bibinfo {author} {\bibfnamefont {Z.}~\bibnamefont
  {{Lin}}}, \bibinfo {author} {\bibfnamefont {D.}~\bibnamefont {{Chen}}},
  \bibinfo {author} {\bibfnamefont {W.}~\bibnamefont {{Lu}}}, \bibinfo {author}
  {\bibfnamefont {X.}~\bibnamefont {{Liang}}}, \bibinfo {author} {\bibfnamefont
  {S.}~\bibnamefont {{Feng}}}, \bibinfo {author} {\bibfnamefont
  {K.}~\bibnamefont {{Yamagami}}}, \bibinfo {author} {\bibfnamefont
  {J.}~\bibnamefont {{Osiecki}}}, \bibinfo {author} {\bibfnamefont
  {M.}~\bibnamefont {{Leandersson}}}, \bibinfo {author} {\bibfnamefont
  {B.}~\bibnamefont {{Thiagarajan}}}, \bibinfo {author} {\bibfnamefont
  {J.}~\bibnamefont {{Liu}}}, \bibinfo {author} {\bibfnamefont
  {C.}~\bibnamefont {{Felser}}},\ and\ \bibinfo {author} {\bibfnamefont
  {J.}~\bibnamefont {{Ma}}},\ }\href@noop {} {\bibinfo {title} {{Observation of
  Giant Spin Splitting and d-wave Spin Texture in Room Temperature Altermagnet
  RuO2}}},\ \Eprint {https://arxiv.org/abs/2402.04995} {arXiv:2402.04995}
  \BibitemShut {NoStop}%
\bibitem [{\citenamefont {Reichlova}\ \emph {et~al.}(2024)\citenamefont
  {Reichlova}, \citenamefont {Lopes~Seeger}, \citenamefont
  {{Gonz{\'a}lez-Hern{\'a}ndez}}, \citenamefont {Kounta}, \citenamefont
  {Schlitz}, \citenamefont {Kriegner}, \citenamefont {Ritzinger}, \citenamefont
  {Lammel}, \citenamefont {Leivisk{\"a}}, \citenamefont {Birk~Hellenes},
  \citenamefont {Olejn{\'i}k}, \citenamefont {Pet{\v r}i{\v c}ek},
  \citenamefont {Dole{\v z}al}, \citenamefont {Horak}, \citenamefont
  {Schmoranzerova}, \citenamefont {Badura}, \citenamefont {Bertaina},
  \citenamefont {Thomas}, \citenamefont {Baltz}, \citenamefont {Michez},
  \citenamefont {Sinova}, \citenamefont {Goennenwein}, \citenamefont
  {Jungwirth},\ and\ \citenamefont {{\v S}mejkal}}]{Reichlova24NCObservation}%
  \BibitemOpen
  \bibfield  {author} {\bibinfo {author} {\bibfnamefont {H.}~\bibnamefont
  {Reichlova}}, \bibinfo {author} {\bibfnamefont {R.}~\bibnamefont
  {Lopes~Seeger}}, \bibinfo {author} {\bibfnamefont {R.}~\bibnamefont
  {{Gonz{\'a}lez-Hern{\'a}ndez}}}, \bibinfo {author} {\bibfnamefont
  {I.}~\bibnamefont {Kounta}}, \bibinfo {author} {\bibfnamefont
  {R.}~\bibnamefont {Schlitz}}, \bibinfo {author} {\bibfnamefont
  {D.}~\bibnamefont {Kriegner}}, \bibinfo {author} {\bibfnamefont
  {P.}~\bibnamefont {Ritzinger}}, \bibinfo {author} {\bibfnamefont
  {M.}~\bibnamefont {Lammel}}, \bibinfo {author} {\bibfnamefont
  {M.}~\bibnamefont {Leivisk{\"a}}}, \bibinfo {author} {\bibfnamefont
  {A.}~\bibnamefont {Birk~Hellenes}}, \bibinfo {author} {\bibfnamefont
  {K.}~\bibnamefont {Olejn{\'i}k}}, \bibinfo {author} {\bibfnamefont
  {V.}~\bibnamefont {Pet{\v r}i{\v c}ek}}, \bibinfo {author} {\bibfnamefont
  {P.}~\bibnamefont {Dole{\v z}al}}, \bibinfo {author} {\bibfnamefont
  {L.}~\bibnamefont {Horak}}, \bibinfo {author} {\bibfnamefont
  {E.}~\bibnamefont {Schmoranzerova}}, \bibinfo {author} {\bibfnamefont
  {A.}~\bibnamefont {Badura}}, \bibinfo {author} {\bibfnamefont
  {S.}~\bibnamefont {Bertaina}}, \bibinfo {author} {\bibfnamefont
  {A.}~\bibnamefont {Thomas}}, \bibinfo {author} {\bibfnamefont
  {V.}~\bibnamefont {Baltz}}, \bibinfo {author} {\bibfnamefont
  {L.}~\bibnamefont {Michez}}, \bibinfo {author} {\bibfnamefont
  {J.}~\bibnamefont {Sinova}}, \bibinfo {author} {\bibfnamefont {S.~T.~B.}\
  \bibnamefont {Goennenwein}}, \bibinfo {author} {\bibfnamefont
  {T.}~\bibnamefont {Jungwirth}},\ and\ \bibinfo {author} {\bibfnamefont
  {L.}~\bibnamefont {{\v S}mejkal}},\ }\href
  {https://doi.org/10.1038/s41467-024-48493-w} {\bibfield  {journal} {\bibinfo
  {journal} {Nat. Commun.}\ }\textbf {\bibinfo {volume} {15}},\ \bibinfo
  {pages} {4961} (\bibinfo {year} {2024})}\BibitemShut {NoStop}%
\bibitem [{\citenamefont {Krempask{\'y}}\ \emph {et~al.}(2024)\citenamefont
  {Krempask{\'y}}, \citenamefont {{\v S}mejkal}, \citenamefont {D'Souza},
  \citenamefont {Hajlaoui}, \citenamefont {Springholz}, \citenamefont
  {Uhl{\'i}{\v r}ov{\'a}}, \citenamefont {Alarab}, \citenamefont
  {Constantinou}, \citenamefont {Strocov}, \citenamefont {Usanov},
  \citenamefont {Pudelko}, \citenamefont {{Gonz{\'a}lez-Hern{\'a}ndez}},
  \citenamefont {Birk~Hellenes}, \citenamefont {Jansa}, \citenamefont
  {Reichlov{\'a}}, \citenamefont {{\v S}ob{\'a}{\v n}}, \citenamefont
  {Gonzalez~Betancourt}, \citenamefont {Wadley}, \citenamefont {Sinova},
  \citenamefont {Kriegner}, \citenamefont {Min{\'a}r}, \citenamefont {Dil},\
  and\ \citenamefont {Jungwirth}}]{Krempasky24NAltermagnetic}%
  \BibitemOpen
  \bibfield  {author} {\bibinfo {author} {\bibfnamefont {J.}~\bibnamefont
  {Krempask{\'y}}}, \bibinfo {author} {\bibfnamefont {L.}~\bibnamefont {{\v
  S}mejkal}}, \bibinfo {author} {\bibfnamefont {S.~W.}\ \bibnamefont
  {D'Souza}}, \bibinfo {author} {\bibfnamefont {M.}~\bibnamefont {Hajlaoui}},
  \bibinfo {author} {\bibfnamefont {G.}~\bibnamefont {Springholz}}, \bibinfo
  {author} {\bibfnamefont {K.}~\bibnamefont {Uhl{\'i}{\v r}ov{\'a}}}, \bibinfo
  {author} {\bibfnamefont {F.}~\bibnamefont {Alarab}}, \bibinfo {author}
  {\bibfnamefont {P.~C.}\ \bibnamefont {Constantinou}}, \bibinfo {author}
  {\bibfnamefont {V.}~\bibnamefont {Strocov}}, \bibinfo {author} {\bibfnamefont
  {D.}~\bibnamefont {Usanov}}, \bibinfo {author} {\bibfnamefont {W.~R.}\
  \bibnamefont {Pudelko}}, \bibinfo {author} {\bibfnamefont {R.}~\bibnamefont
  {{Gonz{\'a}lez-Hern{\'a}ndez}}}, \bibinfo {author} {\bibfnamefont
  {A.}~\bibnamefont {Birk~Hellenes}}, \bibinfo {author} {\bibfnamefont
  {Z.}~\bibnamefont {Jansa}}, \bibinfo {author} {\bibfnamefont
  {H.}~\bibnamefont {Reichlov{\'a}}}, \bibinfo {author} {\bibfnamefont
  {Z.}~\bibnamefont {{\v S}ob{\'a}{\v n}}}, \bibinfo {author} {\bibfnamefont
  {R.~D.}\ \bibnamefont {Gonzalez~Betancourt}}, \bibinfo {author}
  {\bibfnamefont {P.}~\bibnamefont {Wadley}}, \bibinfo {author} {\bibfnamefont
  {J.}~\bibnamefont {Sinova}}, \bibinfo {author} {\bibfnamefont
  {D.}~\bibnamefont {Kriegner}}, \bibinfo {author} {\bibfnamefont
  {J.}~\bibnamefont {Min{\'a}r}}, \bibinfo {author} {\bibfnamefont {J.~H.}\
  \bibnamefont {Dil}},\ and\ \bibinfo {author} {\bibfnamefont {T.}~\bibnamefont
  {Jungwirth}},\ }\href {https://doi.org/10.1038/s41586-023-06907-7} {\bibfield
   {journal} {\bibinfo  {journal} {Nature}\ }\textbf {\bibinfo {volume}
  {626}},\ \bibinfo {pages} {517} (\bibinfo {year} {2024})}\BibitemShut
  {NoStop}%
\bibitem [{\citenamefont {Lee}\ \emph {et~al.}(2024)\citenamefont {Lee},
  \citenamefont {Lee}, \citenamefont {Jung}, \citenamefont {Jung},
  \citenamefont {Kim}, \citenamefont {Lee}, \citenamefont {Seok}, \citenamefont
  {Kim}, \citenamefont {Park}, \citenamefont {{\v S}mejkal}, \citenamefont
  {Kang},\ and\ \citenamefont {Kim}}]{Lee24PRLBroken}%
  \BibitemOpen
  \bibfield  {author} {\bibinfo {author} {\bibfnamefont {S.}~\bibnamefont
  {Lee}}, \bibinfo {author} {\bibfnamefont {S.}~\bibnamefont {Lee}}, \bibinfo
  {author} {\bibfnamefont {S.}~\bibnamefont {Jung}}, \bibinfo {author}
  {\bibfnamefont {J.}~\bibnamefont {Jung}}, \bibinfo {author} {\bibfnamefont
  {D.}~\bibnamefont {Kim}}, \bibinfo {author} {\bibfnamefont {Y.}~\bibnamefont
  {Lee}}, \bibinfo {author} {\bibfnamefont {B.}~\bibnamefont {Seok}}, \bibinfo
  {author} {\bibfnamefont {J.}~\bibnamefont {Kim}}, \bibinfo {author}
  {\bibfnamefont {B.~G.}\ \bibnamefont {Park}}, \bibinfo {author}
  {\bibfnamefont {L.}~\bibnamefont {{\v S}mejkal}}, \bibinfo {author}
  {\bibfnamefont {C.-J.}\ \bibnamefont {Kang}},\ and\ \bibinfo {author}
  {\bibfnamefont {C.}~\bibnamefont {Kim}},\ }\href
  {https://doi.org/10.1103/PhysRevLett.132.036702} {\bibfield  {journal}
  {\bibinfo  {journal} {Phys. Rev. Lett.}\ }\textbf {\bibinfo {volume} {132}},\
  \bibinfo {pages} {036702} (\bibinfo {year} {2024})}\BibitemShut {NoStop}%
\bibitem [{\citenamefont {Osumi}\ \emph {et~al.}(2024)\citenamefont {Osumi},
  \citenamefont {Souma}, \citenamefont {Aoyama}, \citenamefont {Yamauchi},
  \citenamefont {Honma}, \citenamefont {Nakayama}, \citenamefont {Takahashi},
  \citenamefont {Ohgushi},\ and\ \citenamefont {Sato}}]{Osumi24PRBObservation}%
  \BibitemOpen
  \bibfield  {author} {\bibinfo {author} {\bibfnamefont {T.}~\bibnamefont
  {Osumi}}, \bibinfo {author} {\bibfnamefont {S.}~\bibnamefont {Souma}},
  \bibinfo {author} {\bibfnamefont {T.}~\bibnamefont {Aoyama}}, \bibinfo
  {author} {\bibfnamefont {K.}~\bibnamefont {Yamauchi}}, \bibinfo {author}
  {\bibfnamefont {A.}~\bibnamefont {Honma}}, \bibinfo {author} {\bibfnamefont
  {K.}~\bibnamefont {Nakayama}}, \bibinfo {author} {\bibfnamefont
  {T.}~\bibnamefont {Takahashi}}, \bibinfo {author} {\bibfnamefont
  {K.}~\bibnamefont {Ohgushi}},\ and\ \bibinfo {author} {\bibfnamefont
  {T.}~\bibnamefont {Sato}},\ }\href
  {https://doi.org/10.1103/PhysRevB.109.115102} {\bibfield  {journal} {\bibinfo
   {journal} {Phys. Rev. B}\ }\textbf {\bibinfo {volume} {109}},\ \bibinfo
  {pages} {115102} (\bibinfo {year} {2024})}\BibitemShut {NoStop}%
\bibitem [{\citenamefont {Orlova}\ \emph {et~al.}(2024)\citenamefont {Orlova},
  \citenamefont {Avakyants}, \citenamefont {Timonina}, \citenamefont
  {Kolesnikov},\ and\ \citenamefont {Deviatov}}]{Orlova24JLCrossover}%
  \BibitemOpen
  \bibfield  {author} {\bibinfo {author} {\bibfnamefont {N.~N.}\ \bibnamefont
  {Orlova}}, \bibinfo {author} {\bibfnamefont {A.~A.}\ \bibnamefont
  {Avakyants}}, \bibinfo {author} {\bibfnamefont {A.~V.}\ \bibnamefont
  {Timonina}}, \bibinfo {author} {\bibfnamefont {N.~N.}\ \bibnamefont
  {Kolesnikov}},\ and\ \bibinfo {author} {\bibfnamefont {E.~V.}\ \bibnamefont
  {Deviatov}},\ }\href {https://doi.org/10.1134/S0021364024602926} {\bibfield
  {journal} {\bibinfo  {journal} {JETP Lett.}\ }\textbf {\bibinfo {volume}
  {120}},\ \bibinfo {pages} {360} (\bibinfo {year} {2024})}\BibitemShut
  {NoStop}%
\bibitem [{\citenamefont {Reimers}\ \emph {et~al.}(2024)\citenamefont
  {Reimers}, \citenamefont {Odenbreit}, \citenamefont {{\v S}mejkal},
  \citenamefont {Strocov}, \citenamefont {Constantinou}, \citenamefont
  {Hellenes}, \citenamefont {Jaeschke~Ubiergo}, \citenamefont {Campos},
  \citenamefont {Bharadwaj}, \citenamefont {Chakraborty}, \citenamefont
  {Denneulin}, \citenamefont {Shi}, \citenamefont {{Dunin-Borkowski}},
  \citenamefont {Das}, \citenamefont {Kl{\"a}ui}, \citenamefont {Sinova},\ and\
  \citenamefont {Jourdan}}]{Reimers24NCDirect}%
  \BibitemOpen
  \bibfield  {author} {\bibinfo {author} {\bibfnamefont {S.}~\bibnamefont
  {Reimers}}, \bibinfo {author} {\bibfnamefont {L.}~\bibnamefont {Odenbreit}},
  \bibinfo {author} {\bibfnamefont {L.}~\bibnamefont {{\v S}mejkal}}, \bibinfo
  {author} {\bibfnamefont {V.~N.}\ \bibnamefont {Strocov}}, \bibinfo {author}
  {\bibfnamefont {P.}~\bibnamefont {Constantinou}}, \bibinfo {author}
  {\bibfnamefont {A.~B.}\ \bibnamefont {Hellenes}}, \bibinfo {author}
  {\bibfnamefont {R.}~\bibnamefont {Jaeschke~Ubiergo}}, \bibinfo {author}
  {\bibfnamefont {W.~H.}\ \bibnamefont {Campos}}, \bibinfo {author}
  {\bibfnamefont {V.~K.}\ \bibnamefont {Bharadwaj}}, \bibinfo {author}
  {\bibfnamefont {A.}~\bibnamefont {Chakraborty}}, \bibinfo {author}
  {\bibfnamefont {T.}~\bibnamefont {Denneulin}}, \bibinfo {author}
  {\bibfnamefont {W.}~\bibnamefont {Shi}}, \bibinfo {author} {\bibfnamefont
  {R.~E.}\ \bibnamefont {{Dunin-Borkowski}}}, \bibinfo {author} {\bibfnamefont
  {S.}~\bibnamefont {Das}}, \bibinfo {author} {\bibfnamefont {M.}~\bibnamefont
  {Kl{\"a}ui}}, \bibinfo {author} {\bibfnamefont {J.}~\bibnamefont {Sinova}},\
  and\ \bibinfo {author} {\bibfnamefont {M.}~\bibnamefont {Jourdan}},\ }\href
  {https://doi.org/10.1038/s41467-024-46476-5} {\bibfield  {journal} {\bibinfo
  {journal} {Nat. Commun.}\ }\textbf {\bibinfo {volume} {15}},\ \bibinfo
  {pages} {2116} (\bibinfo {year} {2024})}\BibitemShut {NoStop}%
\bibitem [{\citenamefont {Ding}\ \emph {et~al.}(2024)\citenamefont {Ding},
  \citenamefont {Jiang}, \citenamefont {Chen}, \citenamefont {Tao},
  \citenamefont {Liu}, \citenamefont {Li}, \citenamefont {Liu}, \citenamefont
  {Sun}, \citenamefont {Cheng}, \citenamefont {Liu}, \citenamefont {Yang},
  \citenamefont {Zhang}, \citenamefont {Deng}, \citenamefont {Jing},
  \citenamefont {Huang}, \citenamefont {Shi}, \citenamefont {Ye}, \citenamefont
  {Qiao}, \citenamefont {Wang}, \citenamefont {Guo}, \citenamefont {Feng},\
  and\ \citenamefont {Shen}}]{Ding24PRLLarge}%
  \BibitemOpen
  \bibfield  {author} {\bibinfo {author} {\bibfnamefont {J.}~\bibnamefont
  {Ding}}, \bibinfo {author} {\bibfnamefont {Z.}~\bibnamefont {Jiang}},
  \bibinfo {author} {\bibfnamefont {X.}~\bibnamefont {Chen}}, \bibinfo {author}
  {\bibfnamefont {Z.}~\bibnamefont {Tao}}, \bibinfo {author} {\bibfnamefont
  {Z.}~\bibnamefont {Liu}}, \bibinfo {author} {\bibfnamefont {T.}~\bibnamefont
  {Li}}, \bibinfo {author} {\bibfnamefont {J.}~\bibnamefont {Liu}}, \bibinfo
  {author} {\bibfnamefont {J.}~\bibnamefont {Sun}}, \bibinfo {author}
  {\bibfnamefont {J.}~\bibnamefont {Cheng}}, \bibinfo {author} {\bibfnamefont
  {J.}~\bibnamefont {Liu}}, \bibinfo {author} {\bibfnamefont {Y.}~\bibnamefont
  {Yang}}, \bibinfo {author} {\bibfnamefont {R.}~\bibnamefont {Zhang}},
  \bibinfo {author} {\bibfnamefont {L.}~\bibnamefont {Deng}}, \bibinfo {author}
  {\bibfnamefont {W.}~\bibnamefont {Jing}}, \bibinfo {author} {\bibfnamefont
  {Y.}~\bibnamefont {Huang}}, \bibinfo {author} {\bibfnamefont
  {Y.}~\bibnamefont {Shi}}, \bibinfo {author} {\bibfnamefont {M.}~\bibnamefont
  {Ye}}, \bibinfo {author} {\bibfnamefont {S.}~\bibnamefont {Qiao}}, \bibinfo
  {author} {\bibfnamefont {Y.}~\bibnamefont {Wang}}, \bibinfo {author}
  {\bibfnamefont {Y.}~\bibnamefont {Guo}}, \bibinfo {author} {\bibfnamefont
  {D.}~\bibnamefont {Feng}},\ and\ \bibinfo {author} {\bibfnamefont
  {D.}~\bibnamefont {Shen}},\ }\href
  {https://doi.org/10.1103/PhysRevLett.133.206401} {\bibfield  {journal}
  {\bibinfo  {journal} {Phys. Rev. Lett.}\ }\textbf {\bibinfo {volume} {133}},\
  \bibinfo {pages} {206401} (\bibinfo {year} {2024})}\BibitemShut {NoStop}%
\bibitem [{\citenamefont {Yang}\ \emph {et~al.}(2025)\citenamefont {Yang},
  \citenamefont {Li}, \citenamefont {Yang}, \citenamefont {Li}, \citenamefont
  {Zheng}, \citenamefont {Zhu}, \citenamefont {Pan}, \citenamefont {Xu},
  \citenamefont {Cao}, \citenamefont {Zhao}, \citenamefont {Jana},
  \citenamefont {Zhang}, \citenamefont {Ye}, \citenamefont {Song},
  \citenamefont {Hu}, \citenamefont {Yang}, \citenamefont {Fujii},
  \citenamefont {Vobornik}, \citenamefont {Shi}, \citenamefont {Yuan},
  \citenamefont {Zhang}, \citenamefont {Xu},\ and\ \citenamefont
  {Liu}}]{Yang25NCThreedimensional}%
  \BibitemOpen
  \bibfield  {author} {\bibinfo {author} {\bibfnamefont {G.}~\bibnamefont
  {Yang}}, \bibinfo {author} {\bibfnamefont {Z.}~\bibnamefont {Li}}, \bibinfo
  {author} {\bibfnamefont {S.}~\bibnamefont {Yang}}, \bibinfo {author}
  {\bibfnamefont {J.}~\bibnamefont {Li}}, \bibinfo {author} {\bibfnamefont
  {H.}~\bibnamefont {Zheng}}, \bibinfo {author} {\bibfnamefont
  {W.}~\bibnamefont {Zhu}}, \bibinfo {author} {\bibfnamefont {Z.}~\bibnamefont
  {Pan}}, \bibinfo {author} {\bibfnamefont {Y.}~\bibnamefont {Xu}}, \bibinfo
  {author} {\bibfnamefont {S.}~\bibnamefont {Cao}}, \bibinfo {author}
  {\bibfnamefont {W.}~\bibnamefont {Zhao}}, \bibinfo {author} {\bibfnamefont
  {A.}~\bibnamefont {Jana}}, \bibinfo {author} {\bibfnamefont {J.}~\bibnamefont
  {Zhang}}, \bibinfo {author} {\bibfnamefont {M.}~\bibnamefont {Ye}}, \bibinfo
  {author} {\bibfnamefont {Y.}~\bibnamefont {Song}}, \bibinfo {author}
  {\bibfnamefont {L.-H.}\ \bibnamefont {Hu}}, \bibinfo {author} {\bibfnamefont
  {L.}~\bibnamefont {Yang}}, \bibinfo {author} {\bibfnamefont {J.}~\bibnamefont
  {Fujii}}, \bibinfo {author} {\bibfnamefont {I.}~\bibnamefont {Vobornik}},
  \bibinfo {author} {\bibfnamefont {M.}~\bibnamefont {Shi}}, \bibinfo {author}
  {\bibfnamefont {H.}~\bibnamefont {Yuan}}, \bibinfo {author} {\bibfnamefont
  {Y.}~\bibnamefont {Zhang}}, \bibinfo {author} {\bibfnamefont
  {Y.}~\bibnamefont {Xu}},\ and\ \bibinfo {author} {\bibfnamefont
  {Y.}~\bibnamefont {Liu}},\ }\href
  {https://doi.org/10.1038/s41467-025-56647-7} {\bibfield  {journal} {\bibinfo
  {journal} {Nat. Commun.}\ }\textbf {\bibinfo {volume} {16}},\ \bibinfo
  {pages} {1442} (\bibinfo {year} {2025})}\BibitemShut {NoStop}%
\bibitem [{\citenamefont {Liu}\ \emph {et~al.}(2022)\citenamefont {Liu},
  \citenamefont {Li}, \citenamefont {Han}, \citenamefont {Wan},\ and\
  \citenamefont {Liu}}]{Liu22PRXSpinGroup}%
  \BibitemOpen
  \bibfield  {author} {\bibinfo {author} {\bibfnamefont {P.}~\bibnamefont
  {Liu}}, \bibinfo {author} {\bibfnamefont {J.}~\bibnamefont {Li}}, \bibinfo
  {author} {\bibfnamefont {J.}~\bibnamefont {Han}}, \bibinfo {author}
  {\bibfnamefont {X.}~\bibnamefont {Wan}},\ and\ \bibinfo {author}
  {\bibfnamefont {Q.}~\bibnamefont {Liu}},\ }\href
  {https://doi.org/10.1103/PhysRevX.12.021016} {\bibfield  {journal} {\bibinfo
  {journal} {Phys. Rev. X}\ }\textbf {\bibinfo {volume} {12}},\ \bibinfo
  {pages} {021016} (\bibinfo {year} {2022})}\BibitemShut {NoStop}%
\bibitem [{\citenamefont {He}\ \emph {et~al.}(2023)\citenamefont {He},
  \citenamefont {Wang}, \citenamefont {Luo}, \citenamefont {Zeng},
  \citenamefont {Chen},\ and\ \citenamefont {Tang}}]{He23PRLNonrelativistic}%
  \BibitemOpen
  \bibfield  {author} {\bibinfo {author} {\bibfnamefont {R.}~\bibnamefont
  {He}}, \bibinfo {author} {\bibfnamefont {D.}~\bibnamefont {Wang}}, \bibinfo
  {author} {\bibfnamefont {N.}~\bibnamefont {Luo}}, \bibinfo {author}
  {\bibfnamefont {J.}~\bibnamefont {Zeng}}, \bibinfo {author} {\bibfnamefont
  {K.-Q.}\ \bibnamefont {Chen}},\ and\ \bibinfo {author} {\bibfnamefont
  {L.-M.}\ \bibnamefont {Tang}},\ }\href
  {https://doi.org/10.1103/PhysRevLett.130.046401} {\bibfield  {journal}
  {\bibinfo  {journal} {Phys. Rev. Lett.}\ }\textbf {\bibinfo {volume} {130}},\
  \bibinfo {pages} {046401} (\bibinfo {year} {2023})}\BibitemShut {NoStop}%
\bibitem [{\citenamefont {Sheoran}\ and\ \citenamefont
  {Bhattacharya}(2024)}]{Sheoran24PRMNonrelativistic}%
  \BibitemOpen
  \bibfield  {author} {\bibinfo {author} {\bibfnamefont {S.}~\bibnamefont
  {Sheoran}}\ and\ \bibinfo {author} {\bibfnamefont {S.}~\bibnamefont
  {Bhattacharya}},\ }\href {https://doi.org/10.1103/PhysRevMaterials.8.L051401}
  {\bibfield  {journal} {\bibinfo  {journal} {Phys. Rev. Mater.}\ }\textbf
  {\bibinfo {volume} {8}},\ \bibinfo {pages} {L051401} (\bibinfo {year}
  {2024})}\BibitemShut {NoStop}%
\bibitem [{\citenamefont {Liu}\ \emph {et~al.}(2024)\citenamefont {Liu},
  \citenamefont {Yu},\ and\ \citenamefont {Liu}}]{Liu24PRLTwisted}%
  \BibitemOpen
  \bibfield  {author} {\bibinfo {author} {\bibfnamefont {Y.}~\bibnamefont
  {Liu}}, \bibinfo {author} {\bibfnamefont {J.}~\bibnamefont {Yu}},\ and\
  \bibinfo {author} {\bibfnamefont {C.-C.}\ \bibnamefont {Liu}},\ }\href
  {https://doi.org/10.1103/PhysRevLett.133.206702} {\bibfield  {journal}
  {\bibinfo  {journal} {Phys. Rev. Lett.}\ }\textbf {\bibinfo {volume} {133}},\
  \bibinfo {pages} {206702} (\bibinfo {year} {2024})}\BibitemShut {NoStop}%
\bibitem [{\citenamefont {Zeng}\ and\ \citenamefont
  {Zhao}(2024)}]{Zeng24PRBBilayer}%
  \BibitemOpen
  \bibfield  {author} {\bibinfo {author} {\bibfnamefont {S.}~\bibnamefont
  {Zeng}}\ and\ \bibinfo {author} {\bibfnamefont {Y.-J.}\ \bibnamefont
  {Zhao}},\ }\href {https://doi.org/10.1103/PhysRevB.110.174410} {\bibfield
  {journal} {\bibinfo  {journal} {Phys. Rev. B}\ }\textbf {\bibinfo {volume}
  {110}},\ \bibinfo {pages} {174410} (\bibinfo {year} {2024})}\BibitemShut
  {NoStop}%
\bibitem [{\citenamefont {Zhu}\ \emph {et~al.}(2023)\citenamefont {Zhu},
  \citenamefont {Zhuang}, \citenamefont {Wu},\ and\ \citenamefont
  {Yan}}]{Zhu2023Topological}%
  \BibitemOpen
  \bibfield  {author} {\bibinfo {author} {\bibfnamefont {D.}~\bibnamefont
  {Zhu}}, \bibinfo {author} {\bibfnamefont {Z.-Y.}\ \bibnamefont {Zhuang}},
  \bibinfo {author} {\bibfnamefont {Z.}~\bibnamefont {Wu}},\ and\ \bibinfo
  {author} {\bibfnamefont {Z.}~\bibnamefont {Yan}},\ }\href
  {https://doi.org/10.1103/PhysRevB.108.184505} {\bibfield  {journal} {\bibinfo
   {journal} {Phys. Rev. B}\ }\textbf {\bibinfo {volume} {108}},\ \bibinfo
  {pages} {184505} (\bibinfo {year} {2023})}\BibitemShut {NoStop}%
\bibitem [{\citenamefont {Wei}\ \emph {et~al.}(2024)\citenamefont {Wei},
  \citenamefont {Xiang}, \citenamefont {Xu}, \citenamefont {Zhang},
  \citenamefont {Tang},\ and\ \citenamefont {Wang}}]{Wei24PRBGapless}%
  \BibitemOpen
  \bibfield  {author} {\bibinfo {author} {\bibfnamefont {M.}~\bibnamefont
  {Wei}}, \bibinfo {author} {\bibfnamefont {L.}~\bibnamefont {Xiang}}, \bibinfo
  {author} {\bibfnamefont {F.}~\bibnamefont {Xu}}, \bibinfo {author}
  {\bibfnamefont {L.}~\bibnamefont {Zhang}}, \bibinfo {author} {\bibfnamefont
  {G.}~\bibnamefont {Tang}},\ and\ \bibinfo {author} {\bibfnamefont
  {J.}~\bibnamefont {Wang}},\ }\href
  {https://doi.org/10.1103/PhysRevB.109.L201404} {\bibfield  {journal}
  {\bibinfo  {journal} {Phys. Rev. B}\ }\textbf {\bibinfo {volume} {109}},\
  \bibinfo {pages} {L201404} (\bibinfo {year} {2024})}\BibitemShut {NoStop}%
\bibitem [{\citenamefont {Sun}\ \emph {et~al.}(2023)\citenamefont {Sun},
  \citenamefont {Brataas},\ and\ \citenamefont {Linder}}]{Sun23PRBAndreev}%
  \BibitemOpen
  \bibfield  {author} {\bibinfo {author} {\bibfnamefont {C.}~\bibnamefont
  {Sun}}, \bibinfo {author} {\bibfnamefont {A.}~\bibnamefont {Brataas}},\ and\
  \bibinfo {author} {\bibfnamefont {J.}~\bibnamefont {Linder}},\ }\href
  {https://doi.org/10.1103/PhysRevB.108.054511} {\bibfield  {journal} {\bibinfo
   {journal} {Phys. Rev. B}\ }\textbf {\bibinfo {volume} {108}},\ \bibinfo
  {pages} {054511} (\bibinfo {year} {2023})}\BibitemShut {NoStop}%
\bibitem [{\citenamefont {Papaj}(2023)}]{Papaj23PRBAndreev}%
  \BibitemOpen
  \bibfield  {author} {\bibinfo {author} {\bibfnamefont {M.}~\bibnamefont
  {Papaj}},\ }\href {https://doi.org/10.1103/PhysRevB.108.L060508} {\bibfield
  {journal} {\bibinfo  {journal} {Phys. Rev. B}\ }\textbf {\bibinfo {volume}
  {108}},\ \bibinfo {pages} {L060508} (\bibinfo {year} {2023})}\BibitemShut
  {NoStop}%
\bibitem [{\citenamefont {Nagae}\ \emph {et~al.}(2025)\citenamefont {Nagae},
  \citenamefont {Schnyder},\ and\ \citenamefont
  {Ikegaya}}]{Nagae25PRBSpinpolarized}%
  \BibitemOpen
  \bibfield  {author} {\bibinfo {author} {\bibfnamefont {Y.}~\bibnamefont
  {Nagae}}, \bibinfo {author} {\bibfnamefont {A.~P.}\ \bibnamefont
  {Schnyder}},\ and\ \bibinfo {author} {\bibfnamefont {S.}~\bibnamefont
  {Ikegaya}},\ }\href
  {https://journals.aps.org/prb/abstract/10.1103/PhysRevB.111.L100507}
  {\bibfield  {journal} {\bibinfo  {journal} {Phys. Rev. B}\ }\textbf {\bibinfo
  {volume} {111}} (\bibinfo {year} {2025})}\BibitemShut {NoStop}%
\bibitem [{\citenamefont {Zhang}\ \emph
  {et~al.}(2024{\natexlab{a}})\citenamefont {Zhang}, \citenamefont {Hu},\ and\
  \citenamefont {Neupert}}]{Zhang24NCFinitemomentum}%
  \BibitemOpen
  \bibfield  {author} {\bibinfo {author} {\bibfnamefont {S.-B.}\ \bibnamefont
  {Zhang}}, \bibinfo {author} {\bibfnamefont {L.-H.}\ \bibnamefont {Hu}},\ and\
  \bibinfo {author} {\bibfnamefont {T.}~\bibnamefont {Neupert}},\ }\href
  {https://doi.org/10.1038/s41467-024-45951-3} {\bibfield  {journal} {\bibinfo
  {journal} {Nat. Commun.}\ }\textbf {\bibinfo {volume} {15}},\ \bibinfo
  {pages} {1801} (\bibinfo {year} {2024}{\natexlab{a}})}\BibitemShut {NoStop}%
\bibitem [{\citenamefont {Ouassou}\ \emph {et~al.}(2023)\citenamefont
  {Ouassou}, \citenamefont {Brataas},\ and\ \citenamefont
  {Linder}}]{Ouassou23PRLDc}%
  \BibitemOpen
  \bibfield  {author} {\bibinfo {author} {\bibfnamefont {J.~A.}\ \bibnamefont
  {Ouassou}}, \bibinfo {author} {\bibfnamefont {A.}~\bibnamefont {Brataas}},\
  and\ \bibinfo {author} {\bibfnamefont {J.}~\bibnamefont {Linder}},\ }\href
  {https://doi.org/10.1103/PhysRevLett.131.076003} {\bibfield  {journal}
  {\bibinfo  {journal} {Phys. Rev. Lett.}\ }\textbf {\bibinfo {volume} {131}},\
  \bibinfo {pages} {076003} (\bibinfo {year} {2023})}\BibitemShut {NoStop}%
\bibitem [{\citenamefont {Beenakker}\ and\ \citenamefont
  {Vakhtel}(2023)}]{Beenakker23PRBPhaseshifted}%
  \BibitemOpen
  \bibfield  {author} {\bibinfo {author} {\bibfnamefont {C.~W.~J.}\
  \bibnamefont {Beenakker}}\ and\ \bibinfo {author} {\bibfnamefont
  {T.}~\bibnamefont {Vakhtel}},\ }\href
  {https://doi.org/10.1103/PhysRevB.108.075425} {\bibfield  {journal} {\bibinfo
   {journal} {Phys. Rev. B}\ }\textbf {\bibinfo {volume} {108}},\ \bibinfo
  {pages} {075425} (\bibinfo {year} {2023})}\BibitemShut {NoStop}%
\bibitem [{\citenamefont {Sumita}\ \emph {et~al.}(2023)\citenamefont {Sumita},
  \citenamefont {Naka},\ and\ \citenamefont {Seo}}]{Sumita2023Fulde}%
  \BibitemOpen
  \bibfield  {author} {\bibinfo {author} {\bibfnamefont {S.}~\bibnamefont
  {Sumita}}, \bibinfo {author} {\bibfnamefont {M.}~\bibnamefont {Naka}},\ and\
  \bibinfo {author} {\bibfnamefont {H.}~\bibnamefont {Seo}},\ }\href
  {https://doi.org/10.1103/PhysRevResearch.5.043171} {\bibfield  {journal}
  {\bibinfo  {journal} {Phys. Rev. Res.}\ }\textbf {\bibinfo {volume} {5}},\
  \bibinfo {pages} {043171} (\bibinfo {year} {2023})}\BibitemShut {NoStop}%
\bibitem [{\citenamefont {Chakraborty}\ and\ \citenamefont
  {Black-Schaffer}(2024)}]{Chakraborty23Zero}%
  \BibitemOpen
  \bibfield  {author} {\bibinfo {author} {\bibfnamefont {D.}~\bibnamefont
  {Chakraborty}}\ and\ \bibinfo {author} {\bibfnamefont {A.~M.}\ \bibnamefont
  {Black-Schaffer}},\ }\href {https://doi.org/10.1103/PhysRevB.110.L060508}
  {\bibfield  {journal} {\bibinfo  {journal} {Phys. Rev. B}\ }\textbf {\bibinfo
  {volume} {110}},\ \bibinfo {pages} {L060508} (\bibinfo {year}
  {2024})}\BibitemShut {NoStop}%
\bibitem [{\citenamefont {Cheng}\ and\ \citenamefont
  {Sun}(2024)}]{Cheng2024Orientation}%
  \BibitemOpen
  \bibfield  {author} {\bibinfo {author} {\bibfnamefont {Q.}~\bibnamefont
  {Cheng}}\ and\ \bibinfo {author} {\bibfnamefont {Q.}~\bibnamefont {Sun}},\
  }\href {https://doi.org/10.1103/PhysRevB.109.024517} {\bibfield  {journal}
  {\bibinfo  {journal} {Phys. Rev. B}\ }\textbf {\bibinfo {volume} {109}},\
  \bibinfo {pages} {024517} (\bibinfo {year} {2024})}\BibitemShut {NoStop}%
\bibitem [{\citenamefont {Giil}\ and\ \citenamefont
  {Linder}(2024)}]{Giil24PRBSuperconductoraltermagnet}%
  \BibitemOpen
  \bibfield  {author} {\bibinfo {author} {\bibfnamefont {H.~G.}\ \bibnamefont
  {Giil}}\ and\ \bibinfo {author} {\bibfnamefont {J.}~\bibnamefont {Linder}},\
  }\href {https://doi.org/10.1103/PhysRevB.109.134511} {\bibfield  {journal}
  {\bibinfo  {journal} {Phys. Rev. B}\ }\textbf {\bibinfo {volume} {109}},\
  \bibinfo {pages} {134511} (\bibinfo {year} {2024})}\BibitemShut {NoStop}%
\bibitem [{\citenamefont {Lu}\ \emph {et~al.}(2024)\citenamefont {Lu},
  \citenamefont {Maeda}, \citenamefont {Ito}, \citenamefont {Yada},\ and\
  \citenamefont {Tanaka}}]{Lu24PRLphi}%
  \BibitemOpen
  \bibfield  {author} {\bibinfo {author} {\bibfnamefont {B.}~\bibnamefont
  {Lu}}, \bibinfo {author} {\bibfnamefont {K.}~\bibnamefont {Maeda}}, \bibinfo
  {author} {\bibfnamefont {H.}~\bibnamefont {Ito}}, \bibinfo {author}
  {\bibfnamefont {K.}~\bibnamefont {Yada}},\ and\ \bibinfo {author}
  {\bibfnamefont {Y.}~\bibnamefont {Tanaka}},\ }\href
  {https://doi.org/10.1103/PhysRevLett.133.226002} {\bibfield  {journal}
  {\bibinfo  {journal} {Phys. Rev. Lett.}\ }\textbf {\bibinfo {volume} {133}},\
  \bibinfo {pages} {226002} (\bibinfo {year} {2024})}\BibitemShut {NoStop}%
\bibitem [{\citenamefont {Banerjee}\ and\ \citenamefont
  {Scheurer}(2024)}]{Banerjee24PRBAltermagnetic}%
  \BibitemOpen
  \bibfield  {author} {\bibinfo {author} {\bibfnamefont {S.}~\bibnamefont
  {Banerjee}}\ and\ \bibinfo {author} {\bibfnamefont {M.~S.}\ \bibnamefont
  {Scheurer}},\ }\href {https://doi.org/10.1103/PhysRevB.110.024503} {\bibfield
   {journal} {\bibinfo  {journal} {Phys. Rev. B}\ }\textbf {\bibinfo {volume}
  {110}},\ \bibinfo {pages} {024503} (\bibinfo {year} {2024})}\BibitemShut
  {NoStop}%
\bibitem [{\citenamefont {Chakraborty}\ and\ \citenamefont
  {{Black-Schaffer}}()}]{Chakraborty24Perfect}%
  \BibitemOpen
  \bibfield  {author} {\bibinfo {author} {\bibfnamefont {D.}~\bibnamefont
  {Chakraborty}}\ and\ \bibinfo {author} {\bibfnamefont {A.~M.}\ \bibnamefont
  {{Black-Schaffer}}},\ }\href@noop {} {\bibinfo {title} {Perfect
  superconducting diode effect in altermagnets}},\ \Eprint
  {https://arxiv.org/abs/2408.07747} {arXiv:2408.07747} \BibitemShut {NoStop}%
\bibitem [{\citenamefont {Messelot}\ \emph {et~al.}(2024)\citenamefont
  {Messelot}, \citenamefont {Aparicio}, \citenamefont {de~Seze}, \citenamefont
  {Eyraud}, \citenamefont {Coraux}, \citenamefont {Watanabe}, \citenamefont
  {Taniguchi},\ and\ \citenamefont {Renard}}]{Messelot24PRLDirect}%
  \BibitemOpen
  \bibfield  {author} {\bibinfo {author} {\bibfnamefont {S.}~\bibnamefont
  {Messelot}}, \bibinfo {author} {\bibfnamefont {N.}~\bibnamefont {Aparicio}},
  \bibinfo {author} {\bibfnamefont {E.}~\bibnamefont {de~Seze}}, \bibinfo
  {author} {\bibfnamefont {E.}~\bibnamefont {Eyraud}}, \bibinfo {author}
  {\bibfnamefont {J.}~\bibnamefont {Coraux}}, \bibinfo {author} {\bibfnamefont
  {K.}~\bibnamefont {Watanabe}}, \bibinfo {author} {\bibfnamefont
  {T.}~\bibnamefont {Taniguchi}},\ and\ \bibinfo {author} {\bibfnamefont
  {J.}~\bibnamefont {Renard}},\ }\href
  {https://doi.org/10.1103/PhysRevLett.133.106001} {\bibfield  {journal}
  {\bibinfo  {journal} {Phys. Rev. Lett.}\ }\textbf {\bibinfo {volume} {133}},\
  \bibinfo {pages} {106001} (\bibinfo {year} {2024})}\BibitemShut {NoStop}%
\bibitem [{\citenamefont {Furusaki}(1994)}]{Furusaki94PBCMDC}%
  \BibitemOpen
  \bibfield  {author} {\bibinfo {author} {\bibfnamefont {A.}~\bibnamefont
  {Furusaki}},\ }\href {https://doi.org/10.1016/0921-4526(94)90061-2}
  {\bibfield  {journal} {\bibinfo  {journal} {Physica B: Condensed Matter}\
  }\textbf {\bibinfo {volume} {203}},\ \bibinfo {pages} {214} (\bibinfo {year}
  {1994})}\BibitemShut {NoStop}%
\bibitem [{\citenamefont {Asano}(2001)}]{Asano01PRBNumerical}%
  \BibitemOpen
  \bibfield  {author} {\bibinfo {author} {\bibfnamefont {Y.}~\bibnamefont
  {Asano}},\ }\href {https://doi.org/10.1103/PhysRevB.63.052512} {\bibfield
  {journal} {\bibinfo  {journal} {Phys. Rev. B}\ }\textbf {\bibinfo {volume}
  {63}},\ \bibinfo {pages} {052512} (\bibinfo {year} {2001})}\BibitemShut
  {NoStop}%
\bibitem [{\citenamefont {Zhang}\ and\ \citenamefont
  {Trauzettel}(2020)}]{ZhangSB2020PRB}%
  \BibitemOpen
  \bibfield  {author} {\bibinfo {author} {\bibfnamefont {S.-B.}\ \bibnamefont
  {Zhang}}\ and\ \bibinfo {author} {\bibfnamefont {B.}~\bibnamefont
  {Trauzettel}},\ }\href {https://doi.org/10.1103/PhysRevResearch.2.012018}
  {\bibfield  {journal} {\bibinfo  {journal} {Phys. Rev. Res.}\ }\textbf
  {\bibinfo {volume} {2}},\ \bibinfo {pages} {012018} (\bibinfo {year}
  {2020})}\BibitemShut {NoStop}%
\bibitem [{\citenamefont {Golubov}\ \emph {et~al.}(2004)\citenamefont
  {Golubov}, \citenamefont {Kupriyanov},\ and\ \citenamefont
  {Il'ichev}}]{Golubov04RMPCurrentphase}%
  \BibitemOpen
  \bibfield  {author} {\bibinfo {author} {\bibfnamefont {A.~A.}\ \bibnamefont
  {Golubov}}, \bibinfo {author} {\bibfnamefont {M.~Y.}\ \bibnamefont
  {Kupriyanov}},\ and\ \bibinfo {author} {\bibfnamefont {E.}~\bibnamefont
  {Il'ichev}},\ }\href {https://doi.org/10.1103/RevModPhys.76.411} {\bibfield
  {journal} {\bibinfo  {journal} {Rev. Mod. Phys.}\ }\textbf {\bibinfo {volume}
  {76}},\ \bibinfo {pages} {411} (\bibinfo {year} {2004})}\BibitemShut
  {NoStop}%
\bibitem [{\citenamefont {Zhang}\ \emph
  {et~al.}(2024{\natexlab{b}})\citenamefont {Zhang}, \citenamefont {Zarassi},
  \citenamefont {Jarjat}, \citenamefont {{Van de Sande}}, \citenamefont
  {Pendharkar}, \citenamefont {Lee}, \citenamefont {Dempsey}, \citenamefont
  {McFadden}, \citenamefont {Harrington}, \citenamefont {Dong}, \citenamefont
  {Wu}, \citenamefont {Chen}, \citenamefont {Hocevar}, \citenamefont
  {Palmstr{\o}m},\ and\ \citenamefont {Frolov}}]{Zhang24SPLarge}%
  \BibitemOpen
  \bibfield  {author} {\bibinfo {author} {\bibfnamefont {P.}~\bibnamefont
  {Zhang}}, \bibinfo {author} {\bibfnamefont {A.}~\bibnamefont {Zarassi}},
  \bibinfo {author} {\bibfnamefont {L.}~\bibnamefont {Jarjat}}, \bibinfo
  {author} {\bibfnamefont {V.}~\bibnamefont {{Van de Sande}}}, \bibinfo
  {author} {\bibfnamefont {M.}~\bibnamefont {Pendharkar}}, \bibinfo {author}
  {\bibfnamefont {J.~S.}\ \bibnamefont {Lee}}, \bibinfo {author} {\bibfnamefont
  {C.~P.}\ \bibnamefont {Dempsey}}, \bibinfo {author} {\bibfnamefont
  {A.}~\bibnamefont {McFadden}}, \bibinfo {author} {\bibfnamefont {S.~D.}\
  \bibnamefont {Harrington}}, \bibinfo {author} {\bibfnamefont {J.~T.}\
  \bibnamefont {Dong}}, \bibinfo {author} {\bibfnamefont {H.}~\bibnamefont
  {Wu}}, \bibinfo {author} {\bibfnamefont {A.-H.}\ \bibnamefont {Chen}},
  \bibinfo {author} {\bibfnamefont {M.}~\bibnamefont {Hocevar}}, \bibinfo
  {author} {\bibfnamefont {C.~J.}\ \bibnamefont {Palmstr{\o}m}},\ and\ \bibinfo
  {author} {\bibfnamefont {S.~M.}\ \bibnamefont {Frolov}},\ }\href
  {https://doi.org/10.21468/SciPostPhys.16.1.030} {\bibfield  {journal}
  {\bibinfo  {journal} {SciPost Phys.}\ }\textbf {\bibinfo {volume} {16}},\
  \bibinfo {pages} {030} (\bibinfo {year} {2024}{\natexlab{b}})}\BibitemShut
  {NoStop}%
\bibitem [{\citenamefont {Willsch}\ \emph {et~al.}(2024)\citenamefont
  {Willsch}, \citenamefont {Rieger}, \citenamefont {Winkel}, \citenamefont
  {Willsch}, \citenamefont {Dickel}, \citenamefont {Krause}, \citenamefont
  {Ando}, \citenamefont {Lescanne}, \citenamefont {Leghtas}, \citenamefont
  {Bronn}, \citenamefont {Deb}, \citenamefont {Lanes}, \citenamefont {Minev},
  \citenamefont {Dennig}, \citenamefont {Geisert}, \citenamefont {G{\"u}nzler},
  \citenamefont {Ihssen}, \citenamefont {Paluch}, \citenamefont {Reisinger},
  \citenamefont {Hanna}, \citenamefont {Bae}, \citenamefont {Sch{\"u}ffelgen},
  \citenamefont {Gr{\"u}tzmacher}, \citenamefont {{Buimaga-Iarinca}},
  \citenamefont {Morari}, \citenamefont {Wernsdorfer}, \citenamefont
  {DiVincenzo}, \citenamefont {Michielsen}, \citenamefont {Catelani},\ and\
  \citenamefont {Pop}}]{Willsch24NPObservation}%
  \BibitemOpen
  \bibfield  {author} {\bibinfo {author} {\bibfnamefont {D.}~\bibnamefont
  {Willsch}}, \bibinfo {author} {\bibfnamefont {D.}~\bibnamefont {Rieger}},
  \bibinfo {author} {\bibfnamefont {P.}~\bibnamefont {Winkel}}, \bibinfo
  {author} {\bibfnamefont {M.}~\bibnamefont {Willsch}}, \bibinfo {author}
  {\bibfnamefont {C.}~\bibnamefont {Dickel}}, \bibinfo {author} {\bibfnamefont
  {J.}~\bibnamefont {Krause}}, \bibinfo {author} {\bibfnamefont
  {Y.}~\bibnamefont {Ando}}, \bibinfo {author} {\bibfnamefont {R.}~\bibnamefont
  {Lescanne}}, \bibinfo {author} {\bibfnamefont {Z.}~\bibnamefont {Leghtas}},
  \bibinfo {author} {\bibfnamefont {N.~T.}\ \bibnamefont {Bronn}}, \bibinfo
  {author} {\bibfnamefont {P.}~\bibnamefont {Deb}}, \bibinfo {author}
  {\bibfnamefont {O.}~\bibnamefont {Lanes}}, \bibinfo {author} {\bibfnamefont
  {Z.~K.}\ \bibnamefont {Minev}}, \bibinfo {author} {\bibfnamefont
  {B.}~\bibnamefont {Dennig}}, \bibinfo {author} {\bibfnamefont
  {S.}~\bibnamefont {Geisert}}, \bibinfo {author} {\bibfnamefont
  {S.}~\bibnamefont {G{\"u}nzler}}, \bibinfo {author} {\bibfnamefont
  {S.}~\bibnamefont {Ihssen}}, \bibinfo {author} {\bibfnamefont
  {P.}~\bibnamefont {Paluch}}, \bibinfo {author} {\bibfnamefont
  {T.}~\bibnamefont {Reisinger}}, \bibinfo {author} {\bibfnamefont
  {R.}~\bibnamefont {Hanna}}, \bibinfo {author} {\bibfnamefont {J.~H.}\
  \bibnamefont {Bae}}, \bibinfo {author} {\bibfnamefont {P.}~\bibnamefont
  {Sch{\"u}ffelgen}}, \bibinfo {author} {\bibfnamefont {D.}~\bibnamefont
  {Gr{\"u}tzmacher}}, \bibinfo {author} {\bibfnamefont {L.}~\bibnamefont
  {{Buimaga-Iarinca}}}, \bibinfo {author} {\bibfnamefont {C.}~\bibnamefont
  {Morari}}, \bibinfo {author} {\bibfnamefont {W.}~\bibnamefont {Wernsdorfer}},
  \bibinfo {author} {\bibfnamefont {D.~P.}\ \bibnamefont {DiVincenzo}},
  \bibinfo {author} {\bibfnamefont {K.}~\bibnamefont {Michielsen}}, \bibinfo
  {author} {\bibfnamefont {G.}~\bibnamefont {Catelani}},\ and\ \bibinfo
  {author} {\bibfnamefont {I.~M.}\ \bibnamefont {Pop}},\ }\href
  {https://doi.org/10.1038/s41567-024-02400-8} {\bibfield  {journal} {\bibinfo
  {journal} {Nat. Phys.}\ }\textbf {\bibinfo {volume} {20}},\ \bibinfo {pages}
  {815} (\bibinfo {year} {2024})}\BibitemShut {NoStop}%
\bibitem [{\citenamefont {Souto}\ \emph {et~al.}(2022)\citenamefont {Souto},
  \citenamefont {Leijnse},\ and\ \citenamefont {Schrade}}]{Souto22PRL}%
  \BibitemOpen
  \bibfield  {author} {\bibinfo {author} {\bibfnamefont {R.~S.}\ \bibnamefont
  {Souto}}, \bibinfo {author} {\bibfnamefont {M.}~\bibnamefont {Leijnse}},\
  and\ \bibinfo {author} {\bibfnamefont {C.}~\bibnamefont {Schrade}},\ }\href
  {https://doi.org/10.1103/PhysRevLett.129.267702} {\bibfield  {journal}
  {\bibinfo  {journal} {Phys. Rev. Lett.}\ }\textbf {\bibinfo {volume} {129}},\
  \bibinfo {pages} {267702} (\bibinfo {year} {2022})}\BibitemShut {NoStop}%
\bibitem [{\citenamefont {Gingrich}\ \emph {et~al.}(2016)\citenamefont
  {Gingrich}, \citenamefont {Niedzielski}, \citenamefont {Glick}, \citenamefont
  {Wang}, \citenamefont {Miller}, \citenamefont {Loloee}, \citenamefont
  {Pratt~Jr},\ and\ \citenamefont {Birge}}]{Gingrich16NPControllable}%
  \BibitemOpen
  \bibfield  {author} {\bibinfo {author} {\bibfnamefont {E.~C.}\ \bibnamefont
  {Gingrich}}, \bibinfo {author} {\bibfnamefont {B.~M.}\ \bibnamefont
  {Niedzielski}}, \bibinfo {author} {\bibfnamefont {J.~A.}\ \bibnamefont
  {Glick}}, \bibinfo {author} {\bibfnamefont {Y.}~\bibnamefont {Wang}},
  \bibinfo {author} {\bibfnamefont {D.~L.}\ \bibnamefont {Miller}}, \bibinfo
  {author} {\bibfnamefont {R.}~\bibnamefont {Loloee}}, \bibinfo {author}
  {\bibfnamefont {W.~P.}\ \bibnamefont {Pratt~Jr}},\ and\ \bibinfo {author}
  {\bibfnamefont {N.~O.}\ \bibnamefont {Birge}},\ }\href
  {https://doi.org/10.1038/nphys3681} {\bibfield  {journal} {\bibinfo
  {journal} {Nat. Phys}\ }\textbf {\bibinfo {volume} {12}},\ \bibinfo {pages}
  {564} (\bibinfo {year} {2016})}\BibitemShut {NoStop}%
\bibitem [{\citenamefont {Li}\ \emph {et~al.}(2021)\citenamefont {Li},
  \citenamefont {Wang}, \citenamefont {Li}, \citenamefont {Zheng},
  \citenamefont {Brinkman}, \citenamefont {Yu},\ and\ \citenamefont
  {Liao}}]{Li21PRLTopological}%
  \BibitemOpen
  \bibfield  {author} {\bibinfo {author} {\bibfnamefont {C.-Z.}\ \bibnamefont
  {Li}}, \bibinfo {author} {\bibfnamefont {A.-Q.}\ \bibnamefont {Wang}},
  \bibinfo {author} {\bibfnamefont {C.}~\bibnamefont {Li}}, \bibinfo {author}
  {\bibfnamefont {W.-Z.}\ \bibnamefont {Zheng}}, \bibinfo {author}
  {\bibfnamefont {A.}~\bibnamefont {Brinkman}}, \bibinfo {author}
  {\bibfnamefont {D.-P.}\ \bibnamefont {Yu}},\ and\ \bibinfo {author}
  {\bibfnamefont {Z.-M.}\ \bibnamefont {Liao}},\ }\href
  {https://doi.org/10.1103/PhysRevLett.126.027001} {\bibfield  {journal}
  {\bibinfo  {journal} {Phys. Rev. Lett.}\ }\textbf {\bibinfo {volume} {126}},\
  \bibinfo {pages} {027001} (\bibinfo {year} {2021})}\BibitemShut {NoStop}%
\bibitem [{\citenamefont {Mandal}\ \emph {et~al.}(2024)\citenamefont {Mandal},
  \citenamefont {Mondal}, \citenamefont {Stehno}, \citenamefont {Ili{\'c}},
  \citenamefont {Bergeret}, \citenamefont {Klapwijk}, \citenamefont {Gould},\
  and\ \citenamefont {Molenkamp}}]{Mandal24NPMagnetically}%
  \BibitemOpen
  \bibfield  {author} {\bibinfo {author} {\bibfnamefont {P.}~\bibnamefont
  {Mandal}}, \bibinfo {author} {\bibfnamefont {S.}~\bibnamefont {Mondal}},
  \bibinfo {author} {\bibfnamefont {M.~P.}\ \bibnamefont {Stehno}}, \bibinfo
  {author} {\bibfnamefont {S.}~\bibnamefont {Ili{\'c}}}, \bibinfo {author}
  {\bibfnamefont {F.~S.}\ \bibnamefont {Bergeret}}, \bibinfo {author}
  {\bibfnamefont {T.~M.}\ \bibnamefont {Klapwijk}}, \bibinfo {author}
  {\bibfnamefont {C.}~\bibnamefont {Gould}},\ and\ \bibinfo {author}
  {\bibfnamefont {L.~W.}\ \bibnamefont {Molenkamp}},\ }\href
  {https://doi.org/10.1038/s41567-024-02477-1} {\bibfield  {journal} {\bibinfo
  {journal} {Nat. Phys.}\ ,\ \bibinfo {pages} {1}} (\bibinfo {year}
  {2024})}\BibitemShut {NoStop}%
\bibitem [{\citenamefont {Ryazanov}\ \emph {et~al.}(2001)\citenamefont
  {Ryazanov}, \citenamefont {Oboznov}, \citenamefont {Rusanov}, \citenamefont
  {Veretennikov}, \citenamefont {Golubov},\ and\ \citenamefont
  {Aarts}}]{Ryazanov01PRLCoupling}%
  \BibitemOpen
  \bibfield  {author} {\bibinfo {author} {\bibfnamefont {V.~V.}\ \bibnamefont
  {Ryazanov}}, \bibinfo {author} {\bibfnamefont {V.~A.}\ \bibnamefont
  {Oboznov}}, \bibinfo {author} {\bibfnamefont {A.~Y.}\ \bibnamefont
  {Rusanov}}, \bibinfo {author} {\bibfnamefont {A.~V.}\ \bibnamefont
  {Veretennikov}}, \bibinfo {author} {\bibfnamefont {A.~A.}\ \bibnamefont
  {Golubov}},\ and\ \bibinfo {author} {\bibfnamefont {J.}~\bibnamefont
  {Aarts}},\ }\href {https://doi.org/10.1103/PhysRevLett.86.2427} {\bibfield
  {journal} {\bibinfo  {journal} {Phys. Rev. Lett.}\ }\textbf {\bibinfo
  {volume} {86}},\ \bibinfo {pages} {2427} (\bibinfo {year}
  {2001})}\BibitemShut {NoStop}%
\bibitem [{\citenamefont {Bergeret}\ \emph {et~al.}(2001)\citenamefont
  {Bergeret}, \citenamefont {Volkov},\ and\ \citenamefont
  {Efetov}}]{Bergeret01PRLEnhancement}%
  \BibitemOpen
  \bibfield  {author} {\bibinfo {author} {\bibfnamefont {F.~S.}\ \bibnamefont
  {Bergeret}}, \bibinfo {author} {\bibfnamefont {A.~F.}\ \bibnamefont
  {Volkov}},\ and\ \bibinfo {author} {\bibfnamefont {K.~B.}\ \bibnamefont
  {Efetov}},\ }\href {https://doi.org/10.1103/PhysRevLett.86.3140} {\bibfield
  {journal} {\bibinfo  {journal} {Phys. Rev. Lett.}\ }\textbf {\bibinfo
  {volume} {86}},\ \bibinfo {pages} {3140} (\bibinfo {year}
  {2001})}\BibitemShut {NoStop}%
\bibitem [{\citenamefont {Cohen}\ \emph {et~al.}(1962)\citenamefont {Cohen},
  \citenamefont {Falicov},\ and\ \citenamefont
  {Phillips}}]{Cohen62PRLSuperconductive}%
  \BibitemOpen
  \bibfield  {author} {\bibinfo {author} {\bibfnamefont {M.~H.}\ \bibnamefont
  {Cohen}}, \bibinfo {author} {\bibfnamefont {L.~M.}\ \bibnamefont {Falicov}},\
  and\ \bibinfo {author} {\bibfnamefont {J.~C.}\ \bibnamefont {Phillips}},\
  }\href {https://doi.org/10.1103/PhysRevLett.8.316} {\bibfield  {journal}
  {\bibinfo  {journal} {Phys. Rev. Lett.}\ }\textbf {\bibinfo {volume} {8}},\
  \bibinfo {pages} {316} (\bibinfo {year} {1962})}\BibitemShut {NoStop}%
\bibitem [{\citenamefont {Ambegaokar}\ and\ \citenamefont
  {Baratoff}(1963)}]{Ambegaokar63PRLTunneling}%
  \BibitemOpen
  \bibfield  {author} {\bibinfo {author} {\bibfnamefont {V.}~\bibnamefont
  {Ambegaokar}}\ and\ \bibinfo {author} {\bibfnamefont {A.}~\bibnamefont
  {Baratoff}},\ }\href {https://doi.org/10.1103/PhysRevLett.10.486} {\bibfield
  {journal} {\bibinfo  {journal} {Phys. Rev. Lett.}\ }\textbf {\bibinfo
  {volume} {10}},\ \bibinfo {pages} {486} (\bibinfo {year} {1963})}\BibitemShut
  {NoStop}%
\bibitem [{\citenamefont {Abrikosov}\ \emph {et~al.}(2012)\citenamefont
  {Abrikosov}, \citenamefont {Gorkov},\ and\ \citenamefont
  {Dzyaloshinski}}]{Abrikosov12Methods}%
  \BibitemOpen
  \bibfield  {author} {\bibinfo {author} {\bibfnamefont {A.~A.}\ \bibnamefont
  {Abrikosov}}, \bibinfo {author} {\bibfnamefont {L.~P.}\ \bibnamefont
  {Gorkov}},\ and\ \bibinfo {author} {\bibfnamefont {I.~E.}\ \bibnamefont
  {Dzyaloshinski}},\ }\href@noop {} {\emph {\bibinfo {title} {Methods of
  {{Quantum Field Theory}} in {{Statistical Physics}}}}}\ (\bibinfo
  {publisher} {Courier Corporation, New York},\ \bibinfo {year}
  {2012})\BibitemShut {NoStop}%
\end{thebibliography}

\end{document}